\documentclass[prd, aps, twocolumn, nofootinbib, showpacs, superscriptaddress,longbibliography]{revtex4-2}
\usepackage[english]{babel}
\usepackage[utf8]{inputenc}
\usepackage{bm}
\usepackage{subfigure}
\usepackage[colorinlistoftodos, color=green!40, prependcaption]{todonotes}
\usepackage{twoopt}
 
\newcommandtwoopt{\llmet}[2][g][\mu\nu]{#1_{#2}}
\newcommandtwoopt{\uumet}[2][g][\mu\nu]{#1^{#2}}

\newcommand{\vt}{\vartheta}
\newcommand{\eq}[1]{Eq.~\eqref{#1}}
\newcommand{\fig}[1]{Fig.~\ref{#1}}
\newcommand{\tab}[1]{Table~\ref{#1}}

\usepackage{soul}
\usepackage{amsmath, amssymb, amsthm, amsfonts}
\usepackage{mathtools}
\usepackage{physics}
\usepackage{xcolor}
\usepackage{graphicx}
\usepackage{adjustbox}
\usepackage{placeins}
\usepackage[T1]{fontenc}
\usepackage{lipsum}
\usepackage{csquotes}
\usepackage[colorlinks=true,citecolor=green,urlcolor=blue,linkcolor=blue,linktocpage, breaklinks, pdftex, pdftitle={Article}, pdfauthor={Author}]{hyperref}

\usepackage[utf8]{inputenc}
\usepackage{latexsym}
\usepackage{rotating}

\begin{document}
\title{Chaos in Quadratic Gravity}

\author{Alexander Deich}
    \email[Correspondence email address: ]{adeich2@illinois.edu}
    \affiliation{Illinois Center for Advanced Studies of the Universe, Department of Physics, University of Illinois at Urbana-Champaign, Champaign, Illinois, USA}

\author{Alejandro Cárdenas-Avendaño}
\affiliation{Illinois Center for Advanced Studies of the Universe, Department of Physics, University of Illinois at Urbana-Champaign, Champaign, Illinois, USA}
\affiliation{Programa de Matem\'atica, Fundaci\'on Universitaria Konrad Lorenz, 110231 Bogot\'a, Colombia}
\affiliation{Department of Physics, Princeton University, Princeton, NJ, 08544, USA}

\author{Nicolás Yunes}
\affiliation{Illinois Center for Advanced Studies of the Universe, Department of Physics, University of Illinois at Urbana-Champaign, Champaign, Illinois, USA}

\date{\today} 

\begin{abstract}

While recent gravitational wave observations by LIGO and Virgo allow for tests of general relativity in the extreme gravity regime, these observations are still blind to a large swath of phenomena outside these instruments' sensitivity curves. Future gravitational-wave detectors, such as LISA, will enable probes of longer-duration and lower-frequency events. In particular, LISA will enable the characterization of the non-linear dynamics of extreme mass-ratio inspirals, when a small compact object falls into a supermassive black hole. In this paper, we study the motion of test particles around spinning black holes in two quadratic gravity theories: scalar Gauss-Bonnet and dynamical Chern-Simons gravity. We show that geodesic trajectories around slowly rotating black holes in these theories are likely to not have a fourth constant of the motion. In particular, we show that Poincar\'e sections of the orbital phase space present chaotic features that will affect the inspiral of small compact objects into supermassive black holes in these theories. Nevertheless, the characteristic size of these chaotic features is tiny and their location in parameter space is very close to the event horizon of the supermassive black hole. Therefore, the detection of such chaotic features with LISA is likely very challenging, at best.

\end{abstract}

\keywords{first keyword, second keyword, third keyword}


\maketitle

%
%
%
%

\section{Introduction} \label{sec:intro}

Tests of the extreme gravity regime, where gravity is simultaneously strong and dynamical, have become commonplace~\cite{will_2014,yagi2016,berti2018}. The newfound ubiquity of these tests is thanks to the observation of gravitational waves emitted in the inspiral and merger of compact objects by the Laser Interferometer Gravitational-wave Observatory (LIGO) and Virgo. These tests, together with what next-generation detectors such as the space-based Laser Interferometer Space Antenna (LISA)~\cite{Barausse:2020rsu} will enable, allow us to interrogate Einstein's theory of general relativity (GR) with ever finer precision.

While LIGO and Virgo allow for probes of the dynamics of comparable-mass binaries, LISA~\cite{danzmann_2016,amaroseoane2017laser} will allow GR to be tested for an entirely different class of sources. One member of this class are \emph{extreme mass-ratio inspirals} (EMRIs), in which a small black hole (BH) spirals into a supermassive black hole (SMBH)~\cite{Barausse:2020rsu}. EMRIs offer the potential to gain an unique insight into the spacetime geometry of SMBHs because their trajectories, to leading-order in the mass-ratio, are simply geodesics of the SMBH background~\cite{Ryan:1995wh, Glampedakis:2005cf}. This means that EMRIs are particularly sensitive to conserved quantities, such as the orbital energy and orbital angular momentum, as well as to their evolutions~\cite{Glampedakis:2005cf}. The type and number of conserved quantities, in turn, depend on the symmetry structure of the SMBH background. Therefore, EMRI observations have the potential to reveal the symmetries of rotating SMBH spacetimes~\cite{apostolatos2009,Destounis:2021mqv}.

The relation between a set of geodesics, their conserved quantities and the symmetry structure of the SMBH spacetime geometry is highly nontrivial in general~\cite{Babak:2006uv,apostolatos2009}. One method that allows us to establish this relation comes from dynamical systems theory, and in particular, from tools related to the analysis of chaotic systems~\cite{contopoulos_order_2002}.  In Hamiltonian systems (such as a test particle traveling on geodesics in a given spacetime), whether the solution space is chaotic or not depends on its \emph{integrability}~\cite{chaos_review}.
A Hamiltonian system is said to be integrable if there exists a transformation that allows the equations of motion to be written as first integrals of the motion~\cite{contopoulos_order_2002}. This is identical to requiring that there exist at least as many conserved quantities as there are degrees of freedom in the system.  

If either of the above conditions is not met, then the Hamiltonian system is said to be non-integrable, and can permit chaotic motion. This argument can also be reversed:  If one can show that chaotic motion is present in a given system, then one knows the Hamiltonian system is non-integrable, and therefore the number of conserved quantities must be smaller than the degrees of freedom of the system. This, in turn, implies that the spacetime has fewer symmetries than the Kerr solution~\cite{Carter:1968rr}.  In the context of EMRIs~\cite{apostolatos2009, lukes2018, dubeibe2007}, the notion of integrability and how chaotic signatures of the spacetime relate to gravitational-wave observables has attracted much attention~\cite{Destounis:2021mqv}. For example, the abrupt and large changes of the fundamental frequencies of the motion, first reported in Ref.~\cite{apostolatos2009}, have been recently found to translate to ``glitches'' in the GW frequency~\cite{Destounis:2021mqv}. 

The current theoretical understanding of chaotic signatures in EMRIs has been gained by choosing a parametric spacetime that is known to lead to a plethora of chaos~\cite{2011,manko_novikov_2011,Gutierrez-Ruiz:2018tre}. These metrics are typically a solution of the vacuum Einstein equations distinct from the integrable (non-chaotic) Kerr metric~\cite{Carter:1968rr} or designed specifically to be non-integrable~\cite{Destounis:2020kss}. Even though by the no-hair theorems these metrics are expected to be pathological in some way (e.g.~by containing closed timelike curves or lacking a compact event horizon) and may therefore lack astrophysical relevance, these investigations have been extremely informative, as they  allow for a venue to perform tests of GR and study fundamental spacetime symmetries.

But what happens when one studies the possible chaotic features of EMRIs in a better-motivated modified theory of gravity? Recently, Ref.~\cite{cardenas-avendano_exact_2018} investigated whether chaos is present in geodesics of a slowly-rotating BH of dynamical Chern-Simons (dCS) gravity~\cite{alexanderyunes2009}. This parity-violating effective modified gravity theory modifies the Einstein-Hilbert action through a dynamical pseudo-scalar field that couples to the Pontryagin density, and predicts rotating BHs different from the Kerr solution. Given that the solutions to this theory have only been found perturbatively~\cite{ayzenberg_slowly-rotating_2014}, Ref.~\cite{cardenas-avendano_exact_2018} observed that the size of the chaotic features decreased as higher-order expansions were considered in the quadratic sector of the metric. The authors therefore conjectured that these features would disappear altogether given an exact BH metric (i.e.~valid to all orders in spin), implying the existence of a hidden, Carter-like, fourth constant of the motion, associated with an as-yet undiscovered symmetry (in addition to axisymmetry and stationarity). If such a hidden symmetry exists, one expects it is associated with the presence of a Killing tensor~\cite{walker_penrose_1970}. However, recent work had demonstrated that Killing tensors of rank 2, 3, 4, 5 and 6 do not exist for spinning dCS BHs~\cite{owen2021petrov}. This analytical work suggests that chaos may actually be present in geodesic motion around spinning dCS BHs, but its signatures may be so small that they evaded detection until now.  

In this paper, we revisit the question of whether chaos is present in geodesics of spinning BHs in quadratic gravity, including both dCS gravity and scalar Gauss-Bonnet (sGB) gravity~\cite{2009} (a theory similar to dCS but in which a scalar couples to the Kretchmann invariant in the action). We model the spacetime as a resummation of a perturbative solution~\cite{PhysRevD.86.044037}, in which the deformations of the Kerr spacetime are computed in the small-spin and small-coupling approximation, working to fifth order in the former and first order in the latter. We evolve millions of geodesics with an adaptive, $7$--$8$ Runge-Kutta-Verner, custom code that ensures double-precision numerical accuracy over $10^6$ orbits. With these geodesics, we then compute Poincar\'e sections of the orbital phase space, and calculate the rotation curves for geodesics with thousands of initial conditions. From each rotation curve, we then identify non-analytic behavior associated with classical resonances in the phase space, and extract the invariant area in the region of non-analyticity.      

The above analysis, combined with new and faster codes, allows us to analyze the phase space more deeply than ever before, revealing for the first time signatures of chaos in geodesic orbits around both spinning dCS and sGB BHs. We show that these chaotic signatures are robust to the expansion order of the Kerr deformations, as well as to numerical error that can sometimes mimic chaotic structures. The area of the chaotic regions, however, is \textit{exceedingly} small (e.g.,~its width is a millionth of a Schwarzschild radius). The smallness of the chaotic phase space regions implies that their impact in the gravitational waves emitted by EMRIs is likely not detectable. If so, future GW observations that exclude large chaotic signatures as induced by parametric spacetimes would place no constraints on non-Kerr BH solutions that arise in (at least a subset of) actual modified theories of gravity.

At first sight, some of the results described above seem to be in contradiction with those found in Ref.~\cite{cardenas-avendano_exact_2018}. To investigate this further, we reproduced the results in that analysis, and found agreement with their numerical work. However, our new numerical implementation allows for a higher-resolution extraction of chaotic features, and for a deeper exploration of parameter space through parallelization in high-performance computing clusters. These tools yield results that indicate that, although the size of the chaotic regions does decrease with spin order (as found in~\cite{cardenas-avendano_exact_2018}), the change asymptotes to a constant instead of continuing to decrease to zero, (as conjectured in~\cite{cardenas-avendano_exact_2018}). In turn, these results then imply that a fourth constant of the motion does not exist and geodesic motion in these quadratic gravity theories is chaotic. Our results are therefore in agreement with the recent analytical work of Ref.~\cite{owen2021petrov} that proved the non-existence of Killing tensors of rank less than 6 in dCS, and rank 2 in sGB. 

This paper is organized as follows.  Section 2 reviews quadratic gravity and the two specific theories we consider (sGB and dCS gravity).  Section 3 summarizes the mathematical tools we use to detect chaos, discusses the metrics used, and describes some details of the numerical techniques we develop.  Section 4 summarizes the application of these tools to geodesics, in both Kerr and the quadratic gravity theories.  Section 5 concludes and suggests how to take this analysis further in the future.  Appendix A discusses the structure of a metric perturbation necessary to permit chaos in Boyer-Lindquist-like coordinates. Throughout this work we use geometric units in which $G = 1 = c$.

%
%
%
%

\section{Black Holes in Quadratic Gravity}\label{sec:SGB_intro}

In this section, we give a brief description of quadratic gravity, and provide details of the two theories we consider in this paper, sGB and dCS gravity.

\subsection{The quadratic gravity action}
Given the current agreement of GR across several scales and regimes, it may be that modifications to GR appear only in the strong-field regime~\cite{PhysRevD.94.084002}. Modifications can be introduced through a series in higher-than-linear curvature terms in the action, therefore developing an effective field theory. In this context, the Einstein-Hilbert action can be thought of as the leading-order term in such an expansion, and the quadratic correction would be a second-order term~\cite{yunesstein}.  These theories are motivated not only by this effective theory argument, but can also be found arising from  low-energy expansions of certain string theories~\cite{alexanderyunes2009}.  

Quadratic gravity, in particular, refers to a class of effective field theories of modified gravity in which a scalar field couples to quadratic curvature scalars in the action. These theories are defined through the action
\begin{equation}\label{eq:genaction}
    S = S_{\text{EH}} + S_{\text{mat}} + S_\vt + S_{RR},
\end{equation}
where $S_\text{EH}$ denotes the Einstein-Hilbert action, $S_\text{mat}$ is the matter action, $S_\vt$ an action that depends only on the scalar field, and $S_{RR}$ an action that couples the scalar field to a quadratic curvature scalar.

\begin{equation}
    S_\text{EH} = \kappa \int d^4x \, \sqrt{-g} \, R \,,
\end{equation}
with $\kappa = (16\pi G)^{-1}$, $g$ the determinant of the metric tensor and $R=g^{\alpha\beta}g^{\rho\sigma}R_{\rho\alpha\sigma\beta}$ the Ricci scalar, with $R_{\rho\alpha\sigma\beta}$ the Riemann tensor. The action for the scalar field $S_\vt$ is
\begin{equation}
    S_\vt = -\frac{1}{2}\int d^4x \sqrt{-g} \left[\nabla_\mu\vt\nabla^\mu\vt + 2V(\vt)\right],
\end{equation}
where $V(\vt)$ is a potential.  We are here concerned with massless fields, so we set $V(\vt)=0$, leaving only the kinetic piece of $S_\vt$.  To ensure that the theory remains an effective one, we assume $\vartheta$ is small (see e.g.~the discussion in~\cite{2021}). Finally, $S_{RR}$ is the action which couples $\vt$ to some term that is quadratic in the curvature.  While we can imagine constructing many such scalars from curvature invariants, in practice we are here concerned with only two, elaborated upon below.

\subsection{Scalar-Gauss-Bonnet Gravity}

SGB gravity arises from a compactified low-energy expansion of heterotic string theory~\cite{Kanti:1995vq}.  In this effective theory, the action takes the form of \eq{eq:genaction}, with the quadratic term given by~\cite{yagi_challenging_2016}
\begin{equation}
    S_{RR} = \int d^4x \sqrt{-g} \left(\alpha_\text{GB} \, \vt \, \mathcal{G}\right)
\end{equation}
where $\mathcal{G}$ is the Gauss-Bonnet invariant defined as
\begin{equation}
    \mathcal{G} \equiv R^2 - 4 R_{\mu\nu}R^{\mu\nu} + R_{\mu\nu\sigma\rho}R^{\mu\nu\sigma\rho},
\end{equation}
$\vt$ is a scalar field, and $R_{\mu\nu}$ is the Ricci tensor. In geometric units, the constant $\alpha_\text{GB}$ has dimensions of length squared. Observations of gravitational waves emitted by black hole binaries have constrained $\alpha_\text{GB}^{1/2} \leq 5.6\text{km}$ with 90\% confidence  \cite{constraints1}.

The field equations for the theory read~\cite{yagi_challenging_2016}
\begin{align}
    G_{ab} + \frac{\alpha_\text{GB}}{\kappa}\mathcal{D}^{(\vartheta)}_{ab} &= \frac{1}{2\kappa}\left(T^\mathrm{mat.}_{ab} - T^{(\vartheta)}_{ab}\right), 
    \\
    \square \vt &= \alpha_{\text GB} \; {\cal{G}}\,,
\end{align}
where the scalar field stress-energy tensor is
\begin{equation}
    T_{ab}^{(\vt)} = \left[\nabla_a\vt\nabla_b\vt - \frac{1}{2}g_{ab}(\nabla_c\vt\nabla^c\vt-2V(\vt))\right]\,,
    \label{eq-Tab}
\end{equation}
and
\begin{align}
    \mathcal{D}_{ab}^{(\vt)} &\equiv -2R\nabla_a \nabla_b \vt +2(g_{ab}R-2R_{ab})\nabla^c\nabla_c \vt
    \nonumber \\
    &+8R_{c(a}\nabla^c\nabla_{b)}\vt -4g_{ab}R^{cd}\nabla_c \nabla_d \vt
    +4R_{abcd}\nabla^c \nabla^d\vt.
\end{align}

SGB gravity introduces modifications to both spinning (axi-symmetric) and non-spinning (spherically symmetric) BHs~\cite{yagi_challenging_2016}. In all cases, the sGB modifications are proportional to the dimensionless coupling constant
\begin{equation}\label{eq:dcsact}
    \zeta_\text{GB} \equiv \frac{\alpha_\text{GB}^2}{\kappa M^4},
\end{equation}
where $M$ is the black hole mass. Spherically symmetric BHs, however, can easily be shown to lead to integrable (non-chaotic) geodesic orbits (as we review in Appendix A), and thus, we will focus here on spinning BHs.  

\subsection{Dynamical Chern-Simons Gravity}

DCS gravity arises from investigations in string theory~\cite{Alexander:2004xd}, and also through the standard model gravitational anomaly~\cite{dcs2003}, as well as loop quantum gravity~\cite{Taveras:2008yf}. In this effective theory, the action takes the form of \eq{eq:genaction}, with the quadratic term given by~\cite{2009}
\begin{equation}
    S_{RR} = \frac{\alpha_\text{CS}}{4} \int d^4 x \sqrt{-g} \; \tilde{\vt} \; R \tilde{R},
\end{equation}
where the Pontryagin density is
\begin{equation}
    R \tilde{R} \equiv {}^*R^\alpha{}_{\beta}{}{}^{\gamma\delta}R^{\beta}{} _{\alpha\gamma\delta}\,,
\end{equation}
the Riemann tensor's dual is ${}^*R^{\alpha}{}_{\beta}{}^{\gamma\delta} = \frac{1}{2}\epsilon^{\gamma\delta\rho\lambda}R^{\alpha}{}_{\beta\rho\lambda}$, (with $\epsilon^{\mu\nu\alpha\beta}$ the Levi-Civita tensor), $\tilde{\vt}$ is a pseudoscalar field, and $\alpha_\text{CS}$ is a constant with dimensions of length squared in geometric units.  Multi-messenger observations of neutron star have constrained $\alpha_\text{CS}^{1/2} \leq 8.5\text{km}$ with 90\% confidence~\cite{constraints2}.

The field equations for the theory read~\cite{2009}
\begin{align}
    G_{\mu\nu} + \frac{\alpha_\text{CS}}{\kappa}\mathcal{C}^{(\tilde{\vt})}_{\mu\nu} &= \frac{1}{2\kappa}\left(T^\mathrm{mat.}_{\mu\nu} - T^{(\tilde{\vt})}_{\mu\nu}\right), 
    \\
    \square \tilde{\vt} &= -\frac{\alpha_{\text CS}}{4} \; \tilde{R} R\,,
\end{align}
where 
\begin{align}
    \mathcal{C}^{\mu\nu}_{(\tilde{\vt})} &= (\nabla_\sigma\tilde{\vartheta})\epsilon^{\sigma\delta\alpha(\mu}\nabla_\alpha {R^{\nu)}}_\delta + (\nabla_\sigma \nabla_\delta \tilde{\vartheta})\tilde{R}^{\delta(\mu\nu)\sigma},
\end{align}
and $T^{(\tilde{\vt})}_{ab}$ is the stress-energy tensor of the pseudo-scalar field, which is the same as Eq.~\eqref{eq-Tab} but with $\vt \to \tilde{\vt}$.

The action for the non-minimal interaction in \eq{eq:dcsact} introduces modifications from GR only in parity-odd spacetimes, such as for rotating BHs.  As in sGB gravity, dCS modifications to BH spacetimes are proportional to the dimensionless coupling parameter
\begin{equation}\label{eq:dcscoup}
    \zeta_\text{CS} \equiv \frac{\alpha_\text{CS}^2}{\kappa M^4},
\end{equation}
where again $M$ is the BH mass. Unlike sGB gravity, however, dCS modifications are not introduced in spherically symmetric spacetimes~\cite{alexanderyunes2009}.

\subsection{Black hole solutions in sGB and dCS gravity}
Our work relies on the quadratic gravity metrics known analytically from expansions in both slow-spin and small-coupling~\cite{Yagi:2013mbt,ayzenberg_slowly-rotating_2014,Maselli:2017kic}. We will make use of metrics of various orders in spin, and we will even resum these metrics, so that we are able to probe phenomena across a wide range of parameter space in spin and coupling strength.

These metrics were derived following a double approximation scheme as laid out e.g.,~in  \cite{yagi_isolated_2013}.  In this method, two approximations are carried out independently, one in dimensionless spin parameter, $\chi \equiv a/M$ for a BH with dimensional spin parameter $a$ and mass $M$, and one in the coupling term, $\zeta_{\text q}$, where q is the label for the particular theory under consideration.  In both expansions, the parameters are treated as independently small ($\chi \ll 1, \zeta_\text{q}\ll 1$), and thus, the resulting metrics can be thought of as perturbations of a Kerr background metric.  Generically, the GR deformation is expanded to order $(n,m)$ in $(\chi, \zeta_{\text q})$ as follows:
\begin{equation}\label{eq:approx}
    g_{ab} = g_{ab}^{\text Kerr} + \zeta' \sum_{\ell} (\chi')^{\ell} \, \delta g_{ab}^{(\ell)}
\end{equation}
where $\chi'$ and $\zeta'$ are book-keeping parameters that label the expansion order in $\chi$ and $\zeta_{\text q}$, respectively, and $g_{ab}^{\text Kerr}$ is the Kerr metric.  Note that the metric is never expanded beyond \emph{linear} order in $\zeta_\text{q}$, because both sGB and dCS are treated as effective field theories.

Sometimes, we will find it convenient to also expand the Kerr metric in small spin, namely
\begin{equation}\label{eq:Kerrexp}
    g_{ab}^{\text Kerr} = \sum_{k} (\chi')^{\ell} \, g_{ab}^{(k)}\,,
\end{equation}
in which $g_{ab}^{(0)}$ is the Schwarzschild metric. 
 
We adopt the following notation to distinguish these metrics at different expansion orders.  For a metric whose GR sector is expanded to ${\cal{O}}(\chi^n)$ and whose quadratic sector is expanded to ${\cal{O}}(\chi^m\zeta_\text{q})$, we use the notation $\mathcal{O}(\chi^n_\text{GR}, (\chi^m \zeta)_\text{q})$. For example,  a metric labeled $\mathcal{O}(\chi^3_\text{GR}, (\chi^2 \zeta)_\text{dCS})$ is one in which the GR sector is expanded to cubic order in spin and contains up to quadratic orders in the dCS metric deformation. 

All of these expansions present several confounding aspects to the otherwise straightforward search for chaos.  It is already well-established that simply taking a slow-spin expansion to a conventional Kerr metric is sufficient to render the Carter constant unconserved \cite{cardenas-avendano_exact_2018}. In this work, we use metrics that derive from \emph{two} such expansions, one in the GR sector and one in the quadratic sector, both of which could in principle lead to spurious chaotic features. We therefore must be careful to avoid confusing chaotic features that arise due to expansions from true chaotic features that would remain even if we had an exact BH metric.

Another consequence of the approximation scheme described above is the introduction of artificial coordinate singularities. The lowest-order terms of \eq{eq:Kerrexp} are simply the Schwarzschild metric elements, which will contain a coordinate singularity at $r_\text{EH}=2M$, despite the fact that the true location of the event horizon (up to expansion order remainders) is necessarily inside $2M$  \cite{SR_EM,ayzenberg_slowly-rotating_2014}.  We can correct for this through a \emph{resummation} of the metric elements, whereby we perturbatively add terms that correct for the event horizon radius at each subsequent order. These resummed metrics were derived from approximate solutions of  $\mathcal{O}(\chi^2_\text{GR}, (\chi^2 \zeta)_\text{dCS})$ in~\cite{ayzenberg_slowly-rotating_2014}, and we therefore extend them here to approximate metrics of  $\mathcal{O}(\chi^5_\text{GR}, (\chi^5 \zeta)_\text{dCS})$.  The resummed metrics we use here are presented explicitly in~\cite{2022}.

%
%
%
%

\section{Dynamical Systems Theory}\label{sec:theory}

In this section, we introduce some basic concepts from dynamical systems theory that we will employ heavily to study chaos in quadratic gravity. We begin with an introduction to Poincar\'e surfaces and the rotation number as a way to measure chaos. We then present the concept of broken tori and broken symmetries in perturbed spacetimes. We conclude with a discussion of the appropriate geodesics to evolve in order to compute the rotation number. Throughout this section, we draw heavily from the reviews on dynamical systems theory in~\cite{contopoulos_order_2002, chaos_review}.

\subsection{Poincare Surfaces and the Rotation Number}

A Hamiltonian system with $d$ degrees of freedom is said to be ``Liouville integrable'' (or simply integrable) if there exist $d$ constants of the motion which commute  \cite{contopoulos_order_2002}.  For each of these integrable Hamiltonians, there exists a set of action-angle coordinates in which the trajectories lie on hypertori of dimension $d$, embedded in a space of dimension $2d$  \cite{chaos_review}.  If it is ever the case that there are more degrees of freedom than constants of the motion, the system is no longer integrable, and none of the above is necessarily true  \cite{levin2000}. 

\begin{figure*}
    \centering
    \includegraphics[scale=0.4]{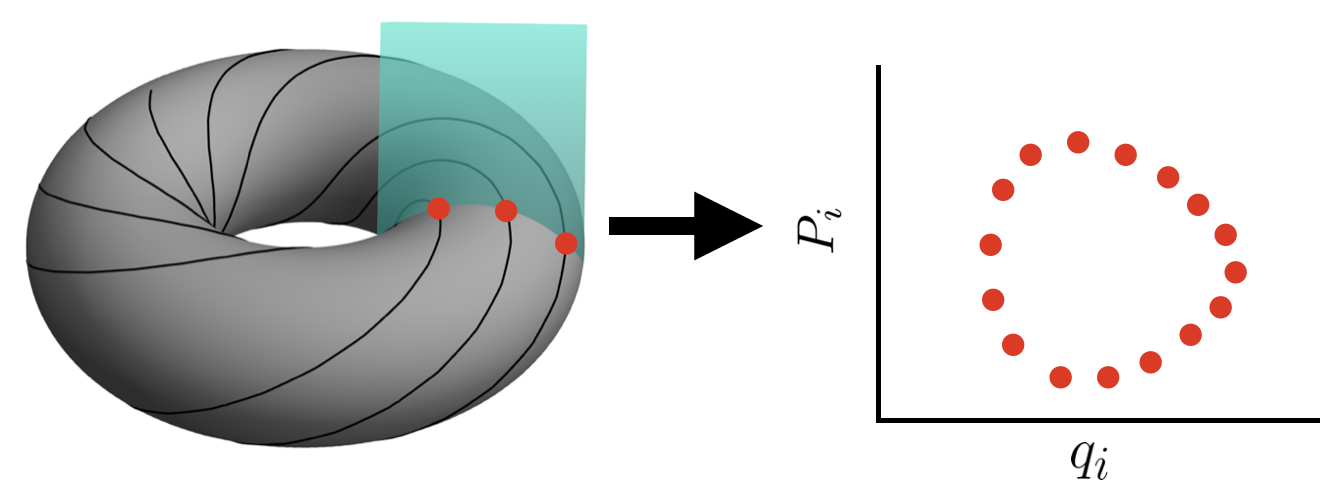}
    \qquad
    \includegraphics[scale=0.4,  trim=2cm 30 0 0]{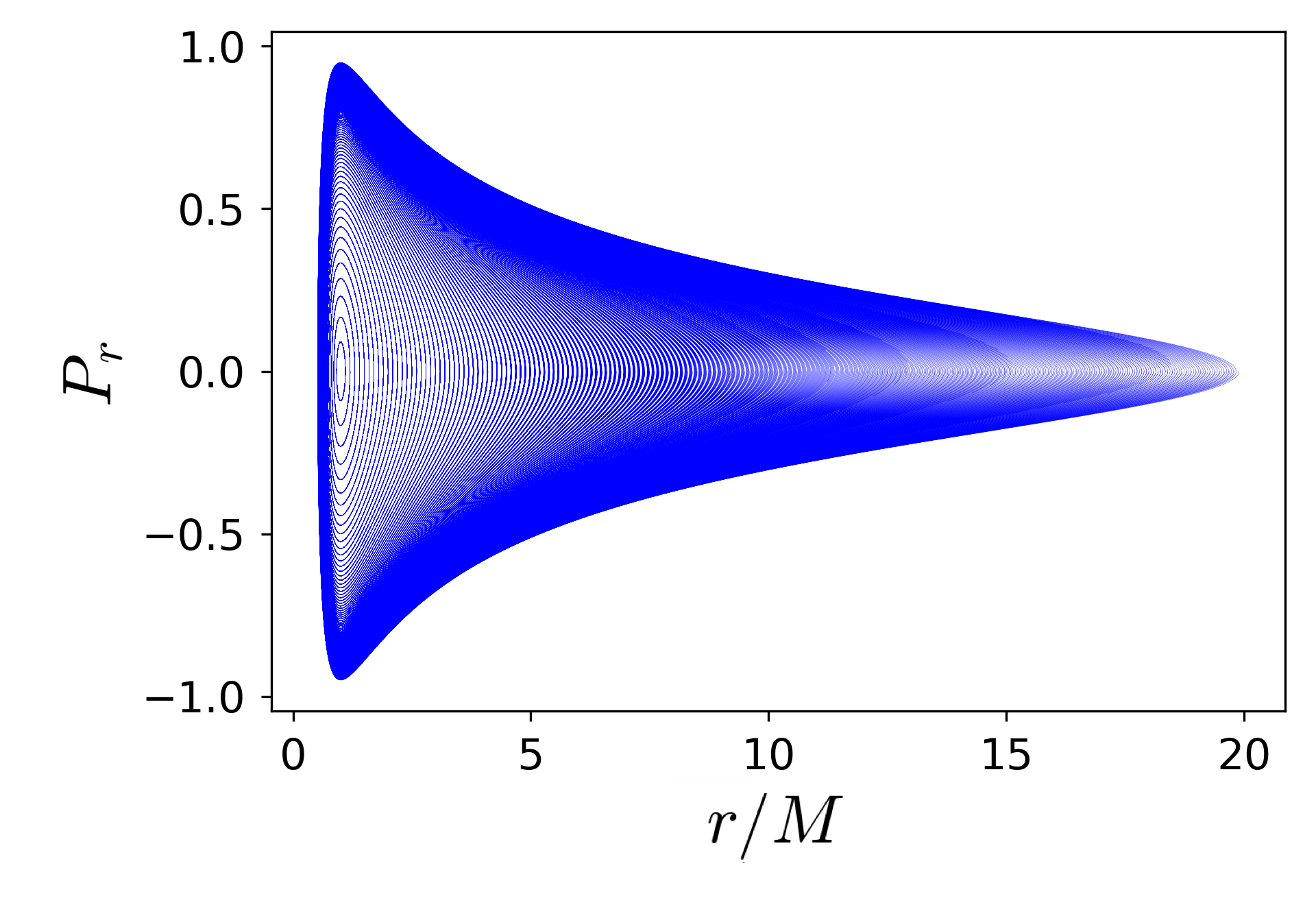}
    \caption{On the left we show an illustration of the torus filled by the trajectories in phase-space and a Poincaré surface. As a trajectory (black line) traverses the torus, the phase-space position is logged whenever the trajectory crosses some plane (cyan rectangle). These positions (red dots) then constitute one Poincaré surface, which are also shown in the phase space of generalized coordinates, as shown in the middle panel.  Depending on how the action-angle coordinates transform, the resulting phase-space cross-section may not be circular. On the right we show several Poincaré surfaces of an integrable system for different trajectories of equal energy and angular momentum.}
    \label{fig:poincare_simple}
\end{figure*}

When dealing with systems of two degrees of freedom, a Poincaré map (also sometimes referred to as a \emph{Poincaré surface of section}) offers a way to quickly understand the behavior of a bundle of trajectories on the surface of a torus.  What would otherwise be a challenging multi-dimensional visualization exercise becomes much simpler to inspect, and several features make themselves much clearer.  To construct a Poincaré map, one integrates a trajectory, and records its phase-space position when it crosses an arbitrary surface (left panel of \fig{fig:poincare_simple}), which we take to be the equatorial plane at $\theta = \pi/2$.

Repeating this procedure for many trajectories across a grid of initial conditions of the same conserved quantities (energy and angular momentum in our case), one arrives at a map that represents a cross-section of a torus (center and right panels of \fig{fig:poincare_simple}), whose dimensions are defined by the trajectory's action-angle coordinates.  As a result of the Kolmogorov-Arnold-Moser (KAM) theorem, if no chaos is present, the curves of the Poincaré map should be nested \cite{contopoulos_order_2002,chaos_review}.  In the presence of chaos, however, the curves start to disintegrate \cite{contopoulos_order_2002}.  The curves branch off into so-called ``Birkhoff islands'', or sometimes explode altogether, leaving a dusty ``sea'' of chaos. The \textit{rotation curve} allows us to quantify the ``amount'' of chaos in each region of the Poincaré map~\cite{contopoulos_order_2002}.

En route to understanding the rotation number, it is useful to consider the following questions: given a phase space trajectory and a slicing of the phase space, how much time elapses between successive crossings of the slice?  Should the crossing frequency be very regular or not predictable?  Given the usual definition of action-angle coordinates (see e.g.~\cite{cardenas-avendano_exact_2018}), one would expect a non-chaotic trajectory to be very regular: action-angle coordinates move with constant-speed motion \cite{chaos_review}. Moreover, the amount by which the trajectory moves between two crossings is completely determined by the ratio $\omega_2/\omega_1$, where $\omega_i$ are the trajectory's angle coordinates at the crossings~\cite{contopoulos_order_2002}. We can identify the orbit that crosses the slicing surface (for us, the surface defined by the equatorial plane $\theta = \pi/2$ with radial momentum $P_r = 0$) at only one point as the \emph{invariant point}, $p_I$. We can then define the angle between subsequent crossings $p_i$ and $p_{i+1}$ via~\cite{contopoulos_order_2002}
\begin{equation}
    \theta_i = \measuredangle\left((p_{i+1} - p_I), (p_i - p_I)\right).
\end{equation}
For an integrable system, it should be the case that $\theta = \omega_2/\omega_1$.  However, the trajectory of a chaotic Hamiltonian cannot be transformed into action-angle coordinates. By the KAM theorem, if your Hamiltonian is a small perturbation from fully integrable, then most trajectories will still be confined to a given torus, and those that are not will not stray very far \cite{contopoulos_order_2002,chaos_review}.  The \emph{rotation number} is therefore defined as~\cite{contopoulos_order_2002}
\begin{equation}\label{eq:rot_num}
    \nu_\theta = \lim_{N\rightarrow \infty}\frac{1}{2\pi N}\sum_i^N \theta_i
\end{equation}
In other words, the rotation number is the average amount by which the trajectory has shifted during one orbit of the torus~\cite{chaos_review}.
\begin{figure*}
    \centering
    \includegraphics[scale=0.5, trim=0 0 0 0]{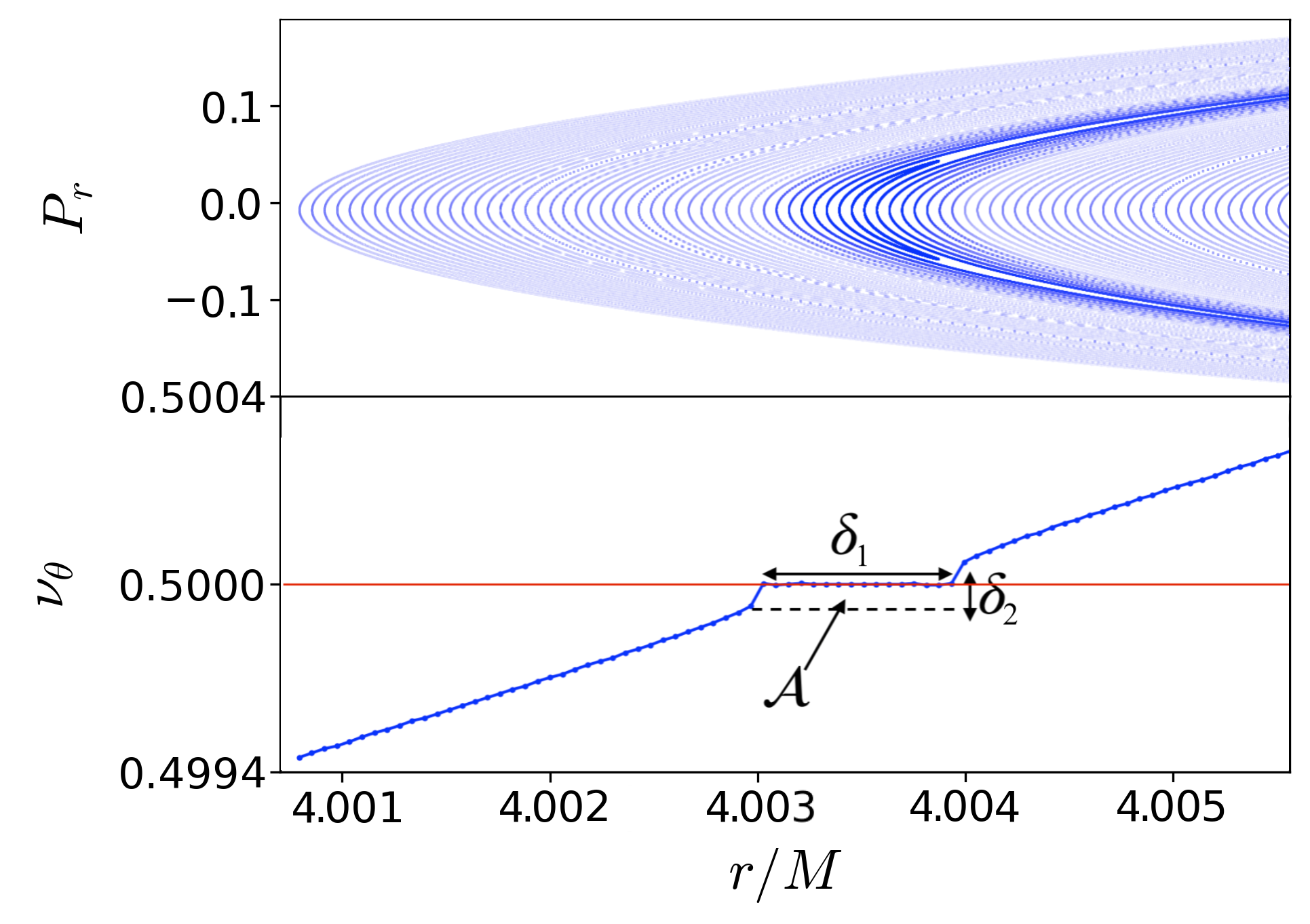}
    \quad
     \includegraphics[scale=0.5]{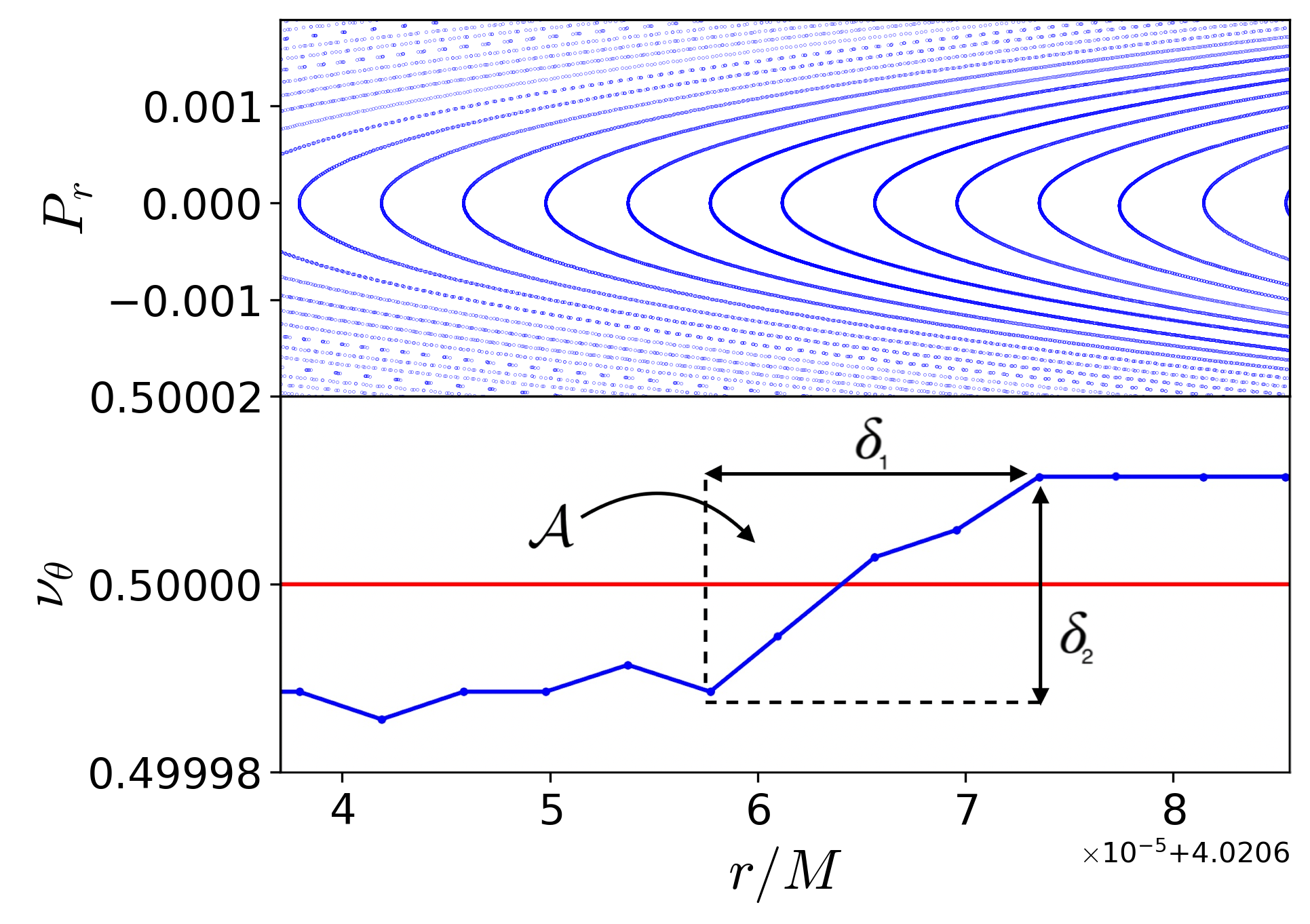}
    \caption{Left: An example of a plateau in the geodesic phase space of an sGB BH at order $\mathcal{O}(\chi^2_\text{GR}, (\chi^5\zeta)_\text{sGB})$, signifying the presence of chaos.  Instead of measuring the width of the plateau, we measure their \textit{area}, $\mathcal{A}$.  This measure allows us to compare plateaus between two chaotic features of different aspect ratios (i.e., where $\delta_1$ has shrunk, so the plateau has become more of a discontinuous jump.) Right: A small plateau, found in the geodesic phase space of an sGB BH expanded to $\mathcal{O}(\chi^7_\text{GR}, (\chi^5\zeta)_\text{sGB})$. For very small plateaus, $\delta_1$ can shrink more than $\delta_2$, changing the aspect ratio.  Measuring $\delta_1$ alone can indicate, incorrectly, that the plateau is vanishing. The plateau area, on the other hand, allows us to compare plateaus of different aspect ratios more meaningfully.  In both panels, $E = 0.995\mu$, $L = 3.75365\mu M$ and $\chi = 0.2$.}
    \label{fig:a_def}
\end{figure*}
By calculating the average angle between successive crossings of the chosen slice, we can compare the ``amount'' of chaos of one trajectory to another.  In particular, if the rotation number either ``flattens out'' into a so-called ``plateau'' (left panel of \fig{fig:a_def}) or suddenly makes a discontinuous jump (right panel of \fig{fig:a_def}), that is a tell-tale sign of chaos. We will refer to all such signs as non-analytic features of the rotation curve.

\subsection{Measuring the Broken Tori}

When the perturbations from the integrable system are small, as the ones considered in this work, searching for regions of broken toroidal structure in phase-space is computationally expensive.  In addition to being a very large space, any detection must be shown as arising due to the equations of motion themselves, and not any artifact of numerical precision.  In order to reduce the region of phase space being searched over, we need a hint about where to start looking.  Fortunately, it can be shown~\cite{contopoulos_order_2002,chaos_review} that, as a result of the Poincar\'e-Birkhoff theorem~\cite{contopoulos_order_2002}, trajectories of a perturbed system are most likely to exhibit chaos if they lie near trajectories with integer ratio of frequencies $\omega$.  Such frequency ratios are called \emph{resonant}.  For this reason, we look first to the $1/2$ and $2/3$ resonances for any signatures of chaos.

Once non-analytic features in the rotation curve are found, we must also find a way to quantify the size of the chaotic region. One way to do so is through the area $\mathcal{A}$ of the region in phase space in which these non-analytic features are contained. For example, let us consider plateaus. As the size of the plateau decreases, the plateau's aspect ratio quickly flips, becoming larger in the $P_r$ axis than in the $r$ axis.  While past work has measured only the width of the plateau, this loses meaning as the plateau shrinks.  Therefore, it is easier to compare plateau sizes if we concern ourselves with the area $\mathcal{A}$ of the plateau, defined via
\begin{align}
    {\cal{A}} = \int_{\nu_{\text{min}}}^{\nu_{\text{max}}} \int_{r_\text{min}}^{r_\text{max}} d\nu \; dr\,,
\end{align}
where $r_{\text{min,max}}$ and $\nu_{\text{min,max}}$ are the radii and rotation numbers at which the non-analytic features first appear and disappear as we sweep through initial conditions.  For the non-analytic features we consider, $\mathcal{A}$ can be well approximated as a rectangle, which is then given by the width, $\delta_1$ of the plateau times its height, $\delta_2$ (\fig{fig:a_def}), i.e.,~${\cal{A}} = \delta_1 \delta_2$.

For systems without dissipation, like the one we are studying here, the resulting phase space portrait is independent of the initial conditions. However, the measure we described above depends on where in the phase portrait it is implemented. For this work, we are only concerned with the 1/2-resonance, which has a Birkhoff island whose maximum width is at $P_r = 0$, and that is why we see plateaus for some cases. If one does not use points in the phase space with $P_r = 0$, the rotation number for this same resonance may not show a plateau, and the described area will be even smaller. Nevertheless, it can still be used and compared to other conditions if the phase portrait is consistently measured, i.e., in the same way for all the analyzed cases. Thus, for different resonances, one may need to explore the phase portrait and find where these effects may be maximized.

On the other hand, if one includes dissipation, the initial conditions will impact the dynamics of the particle~\cite{2010,Bronicki:2022eqa,PhysRevD.104.064023}, as it will cross different structures in phase space differently. As we are not considering a particle crossing these structures, we just need to focus on the depicted character of the phase portrait to study the integrability of the aforementioned quadratic theories of gravity.

\subsection{Broken Symmetries in Perturbed Spacetimes}
The KAM theorem suggests at least two ways in which a non-integrable Hamiltonian can be perturbed to lead to chaos: either reduce the number of conserved quantities, or increase the number of degrees of freedom. Recall that the Hamiltonian for geodesic motion on arbitrary backgrounds is
\begin{equation}\label{eq:partham}
    H = \frac{1}{2\mu}g^{\mu\nu}P_\mu P_\nu,
\end{equation}
for a trajectory with four-momentum $P_\mu = \mu \, u_\mu$ of a test particle with mass $\mu$ and four-velocity $u_\mu$. In our case, the metric $g_{\mu \nu}$ is given by Eq.~\eqref{eq:approx}, which can be thought of as a perturbation of the Kerr spacetime. The symmetries of the Kerr metric disallow any chaotic behavior in geodesic trajectories~\cite{Carter:1968rr}. Perturbations to this Hamiltonian introduced by the deformations of the Kerr metric can then lead to chaos either by changing the number of conserved quantities or degrees of freedom. We know the number of degrees of freedom remains fixed (there are still 4 spacetime dimensions in sGB and dCS gravity), but different theories may permit different numbers of conserved quantities. Therefore, if chaos is found in sGB or dCS gravity, then it must be that the number of conserved quantities has been reduced relative to GR. 

A particle traveling along a geodesic in a Kerr spacetime possesses 4 conserved quantities: the angular momentum in the $z-$axis $L$, the total energy $E$, the rest-mass of the particle $\mu$ and the Carter constant $C$~\cite{Carter:1968rr}.  The conserved quantities $E$ and $L$ are defined from the contractions $E = -\mu^{-1} \xi^\mu_{(t)}P_\mu$ and $L = \mu^{-1} \xi^\mu_{(\phi)}P_\mu$, where $\xi^\mu_{(X)}$ is the Killing vector associated with the $X$ coordinate.  The Carter constant is derived from a second-rank Killing tensor. The angular moment and the total energy arise from the existence of azimuthal and time-like Killing vectors, associated with axisymmetry and stationarity. The conservation of $\mu$ follows from the conservation of the metric signature upon geodesic evolution. Therefore, these three conserved quantities always exist in geodesics around stationary and axisymmetric backgrounds. A Killing tensor, however, need not exist in general for modified theories.  Determining the nonexistence of Killing tensors of arbitrary rank is generally very difficult.  While searches have been performed recently for dCS and sGB gravity, finding that Killing tensors do not exist for slowly-rotating BH spacetimes up to rank 6~\cite{owen2021petrov}, the existence of higher-rank Killing tensors cannot be ruled out.

We will here adopt an alternative path to characterize the number of conserved quantities of a given theory:  given a modified (stationary and axisymmetric) BH spacetime, evolve many test particles along geodesics to look for signatures of chaos in their associated Poincaré surfaces~\cite{1997, 2010}.  This will provide an indication of whether there is chaos in the spacetime (to within the numerical precision of our integrator\footnote{Indeed, for the purposes of setting bounds on coupling parameters, this is all that is needed, as machine precision currently much outpaces even the future detectors' ability to constrain modified theories.}). The aforementioned procedure has been applied to various spacetimes to study their integrability properties (see, e.g., Refs.~\cite{1997,Kiuchi:2004bv,Gair:2007kr,apostolatos2009,cardenas-avendano_exact_2018,Gutierrez-Ruiz:2018tre,Zelenka:2017aqn}).

We are now equipped to formulate our original question differently:  Do the metrics of quadratic gravity break any symmetries of Kerr, and if they do, how significant are the resulting features of chaos?  Ultimately, these features are what will allow any sort of observational constraint to be placed on the coupling constant $\alpha_{\text{GB},\text{CS}}$.

%
%
%
%

\section{Geodesics in Quadratic Gravity}

Having laid the foundations for the modified gravity theories we will investigate, as well as some basics of dynamical systems theory, we can now proceed to investigate whether geodesics present chaotic features in quadratic gravity.

\subsection{Choosing Appropriate Geodesics and Initial Conditions}

What kinds of particle orbits are most useful for carrying out the analysis described above?  While we can imagine doing these calculations for any kind of geodesic, Poincaré sections are relatively data-intensive, so we desire geodesics that can provide a theoretically infinite number of surface crossings.  We therefore restrict ourselves to bound orbits\footnote{Other studies have also investigated plunging orbits, but these revealed that the resulting features are relatively small~\cite{cardenas-avendano_exact_2018}.}. In addition, we should also focus on those geodesics that approach the BH event horizon as close as possible, to ensure the geodesics sample as strong a gravitational field as possible. Quadratic gravity modifications naturally become stronger the larger the spacetime curvature, and therefore any chaotic features due to the modification will scale with the distance to the BH event horizon. 

\begin{figure*}
  \centering
  \subfigure{\includegraphics[scale=0.47, trim=0 0 0 0]{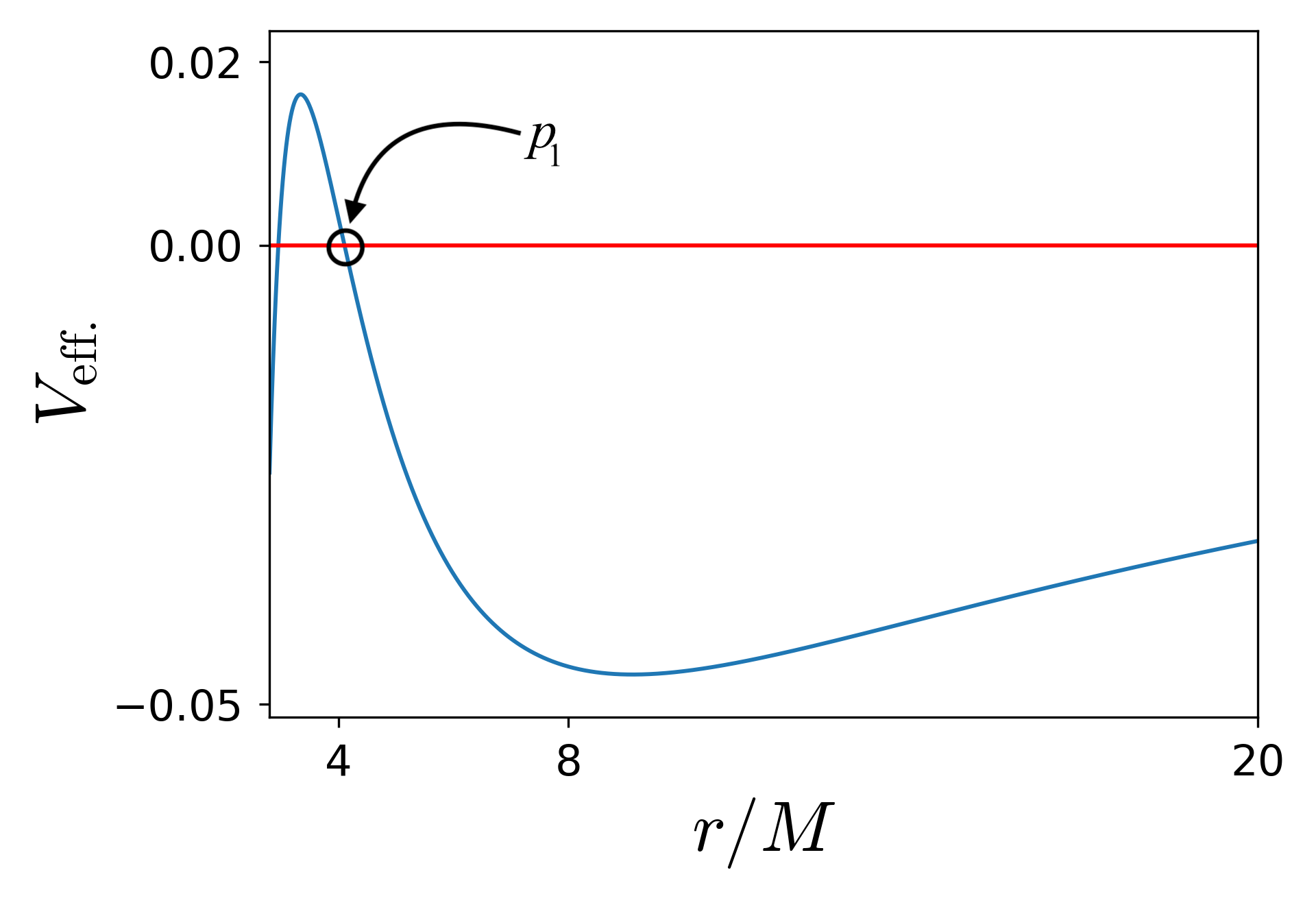}}
  \subfigure{\includegraphics[scale=0.2]{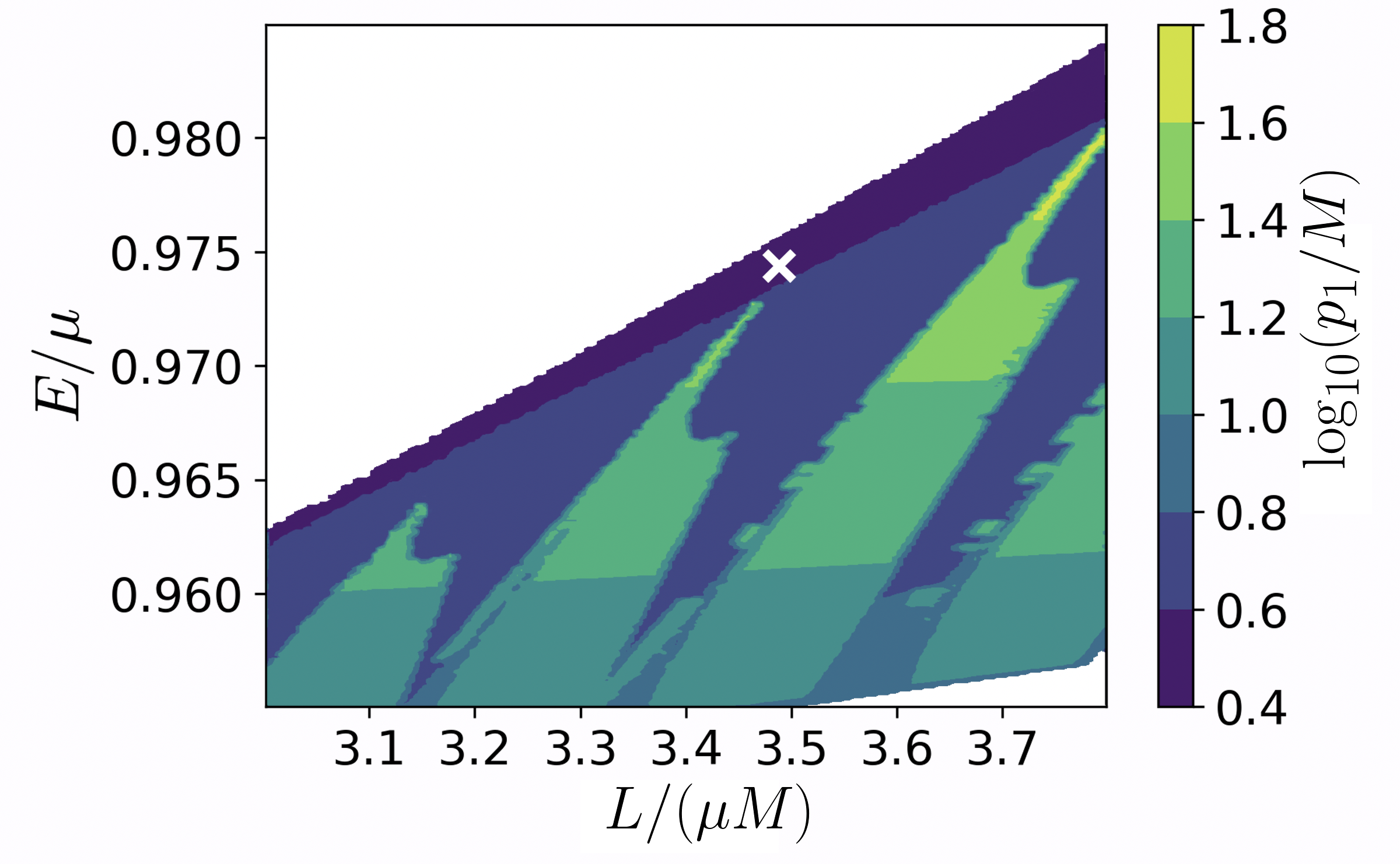}}\quad
  \caption{Left: Effective potential for a given choice of $(E,L)$ and a small value for its second root $p_1/M$. Note that $p_1/M$ must exist in order to achieve a bound orbit, but at the same time, $p_1$ must be minimized in order to probe the strongest field possible. Right: Values of the effective potential's second root, $p_1/M$ for which values of specific energy and angular momentum lead to bound geodesics. This contour heat map helps determine the initial conditions that will allow for the smallest values of $p_1/M$. For this calculation, we use a Kerr metric and $\chi = 0.3$. The values of $E$ and $L$ we use in Fig. 5, $E=0.97406\mu,L = 3.49916\mu M$, are indicated with a white `$\cross$'.  }
  \label{fig:veff_minimization}
\end{figure*}

The region of phase space that a particle explores is controlled by its effective potential.  In a geometric theory of gravity, one can reduce the Hamiltonian for particle motion in a stationary and axisymmetric background to two degrees of freedom via the normalization condition $P_\alpha P^\alpha = -\mu^2$, where recall that $P_\alpha$ is the particle's     4-momentum. We can then rewrite \eq{eq:partham} as 
\begin{equation}
\frac{1}{2} P_r^2 + \frac{1}{2} P_{\theta}^2 = V_\text{eff.}\,,    
\label{eq-H-Veff}
\end{equation} 
where we have defined the effective potential
\begin{equation}
    V_\text{eff.} = \frac{1}{2}\left(\frac{g_{\phi\phi} E^2 + 2g_{t\phi} E L+L^2g_{tt}}{g_{tt}g_{\phi\phi} - g_{t\phi}^2}   +\mu^2 \right),
\end{equation}
and recall that $E$ and $L$ are the particle's energy and $z-$angular momentum, respectively. If for some choice of $(E,L)$, the effective potential contains a local minimum, then geodesics with those choices of $(E,L)$ are bounded. For example, when we plot $V_\text{eff.}$ for a bound system, we find that it presents a distinct local minimum, recognizable from Newtonian dynamics, as shown in \fig{fig:veff_minimization}.  

To amplify the effects of chaos, we must now choose the subset of $(E,L)$ that not only lead to bound geodesic motion, but also allow these geodesics to sample the largest possible spacetime curvatures. The latter is accomplished by requiring that the geodesics explore regimes of spacetime close to the BH event horizon. For this to occur, we must then require that the second root of $V_\text{eff.}=0$ (labeled $p_1$ in \fig{fig:veff_minimization}) is as close as possible to the BH event horizon.

We therefore formulate this as a minimization problem in the two-dimensional space of possible values for $E$ and $L$.  To perform the minimization of the location of the second root, we grid in $E$ and $L$ and find those combinations that permit a local minimum in the effective potential.  Then, by a root finding algorithm we find the set of parameters that minimizes the second root of the effective potential, and we use this root as the left-most initial radius for the geodesic runs.  To perform this minimization, we found it was sufficient to set the numerical precision to $10^{-10}$. The value of the second root as a function of $E$ and $L$ can be seen in \fig{fig:veff_minimization}.

\begin{figure*}
  \centering
  \subfigure{\includegraphics[scale=0.55]{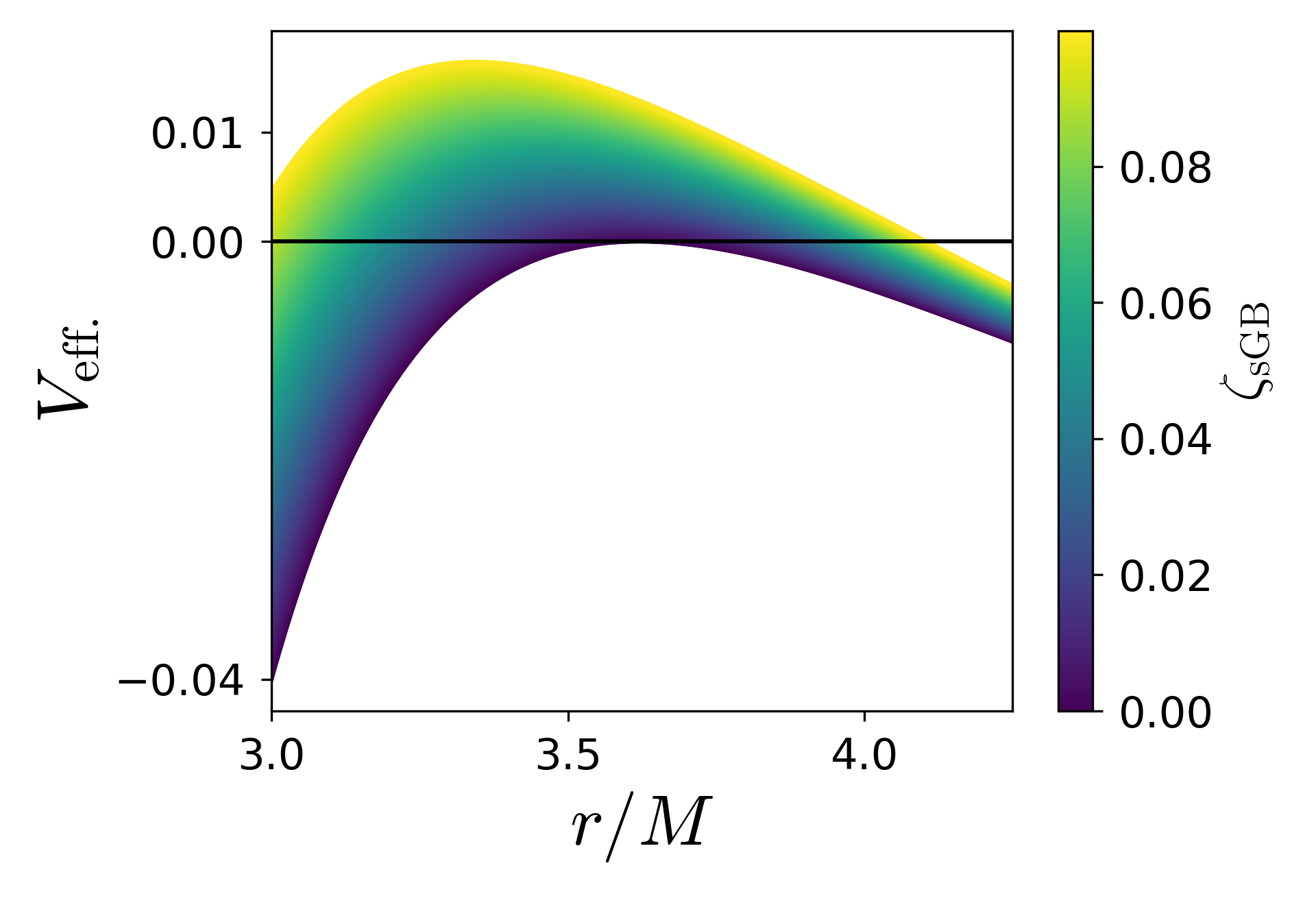}}\quad
  \subfigure{\includegraphics[scale=0.525, trim=0 -.6cm 0 0]{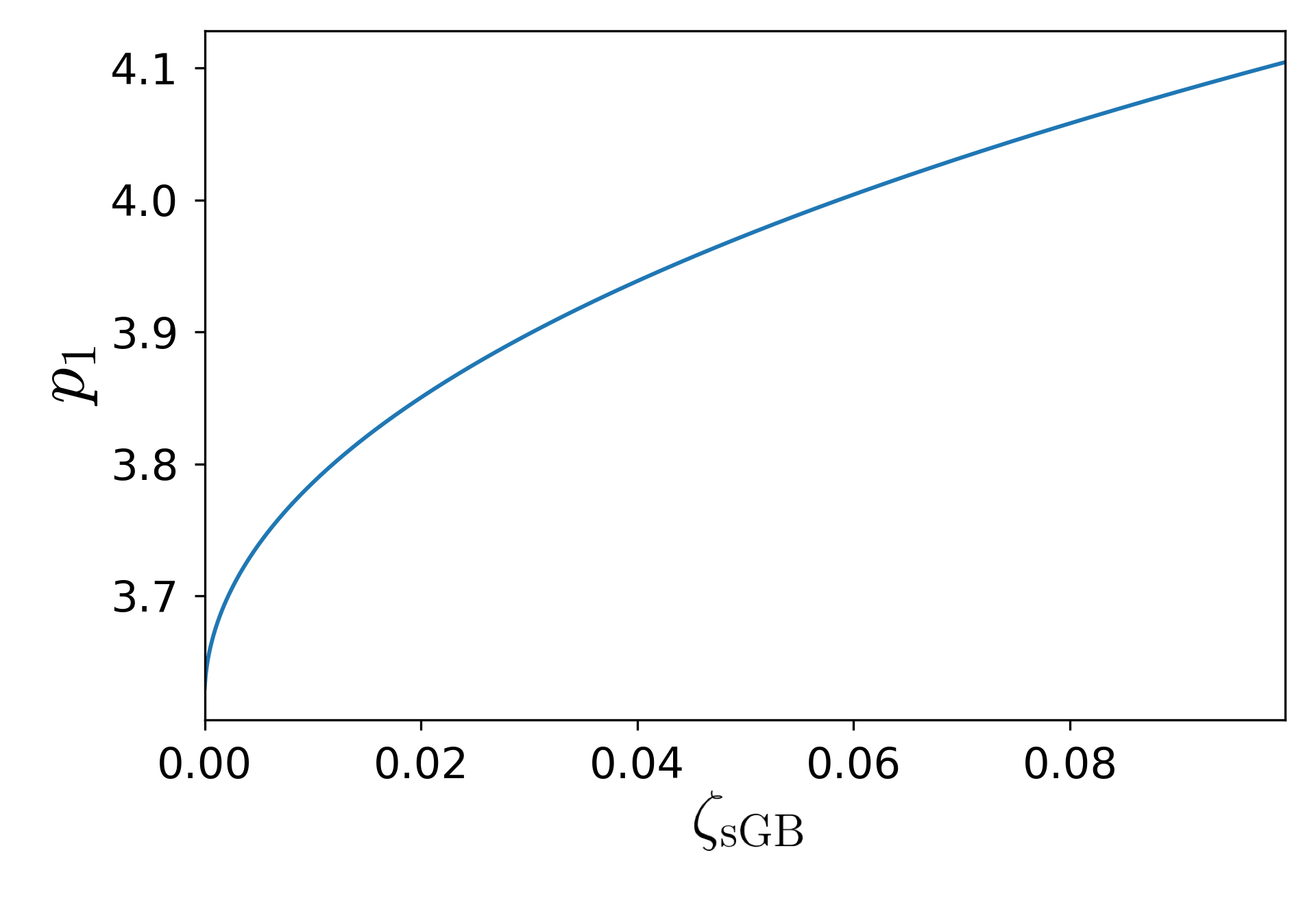}}
  \caption{Effective potentials with $E = 0.995\mu$, $L = 3.75365\mu M$ and $\chi = 0.2$ for different values of the coupling parameter $\zeta_\text{sGB}$ show that the innermost radius accessible to bound orbits, $p_1$, increases monotonically with $\zeta_\text{sGB}$ (right). Thus, we expect any effect that depends on the modification to decrease with the strength of the coupling parameter. The same relationship also holds in dCS gravity.}
  \label{fig:veff_zeta}
\end{figure*}

\begin{figure*}
  \centering
  \subfigure{\includegraphics[scale=0.55]{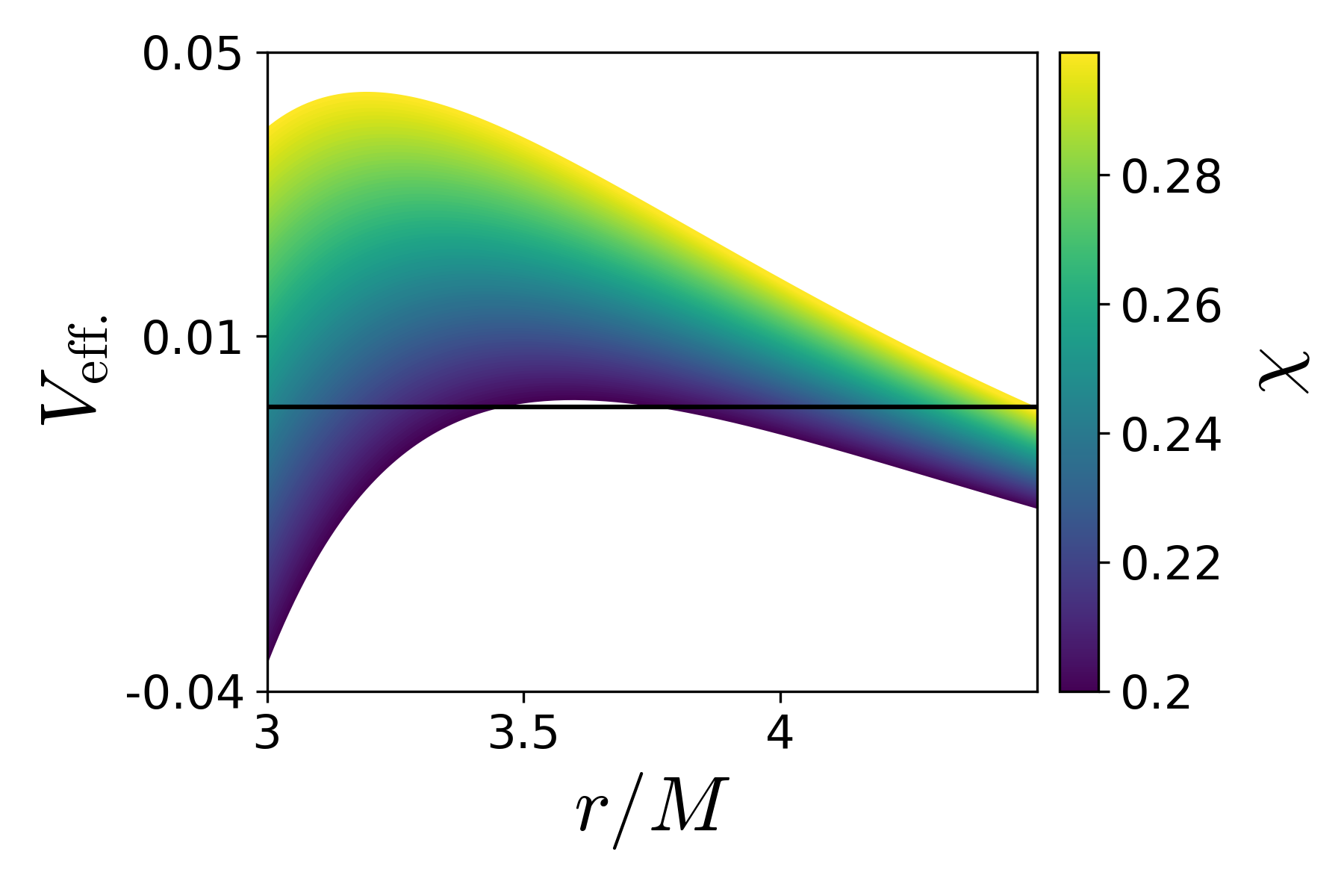}}\quad
  \subfigure{\includegraphics[scale=0.525, trim=0 -.6cm 0 0]{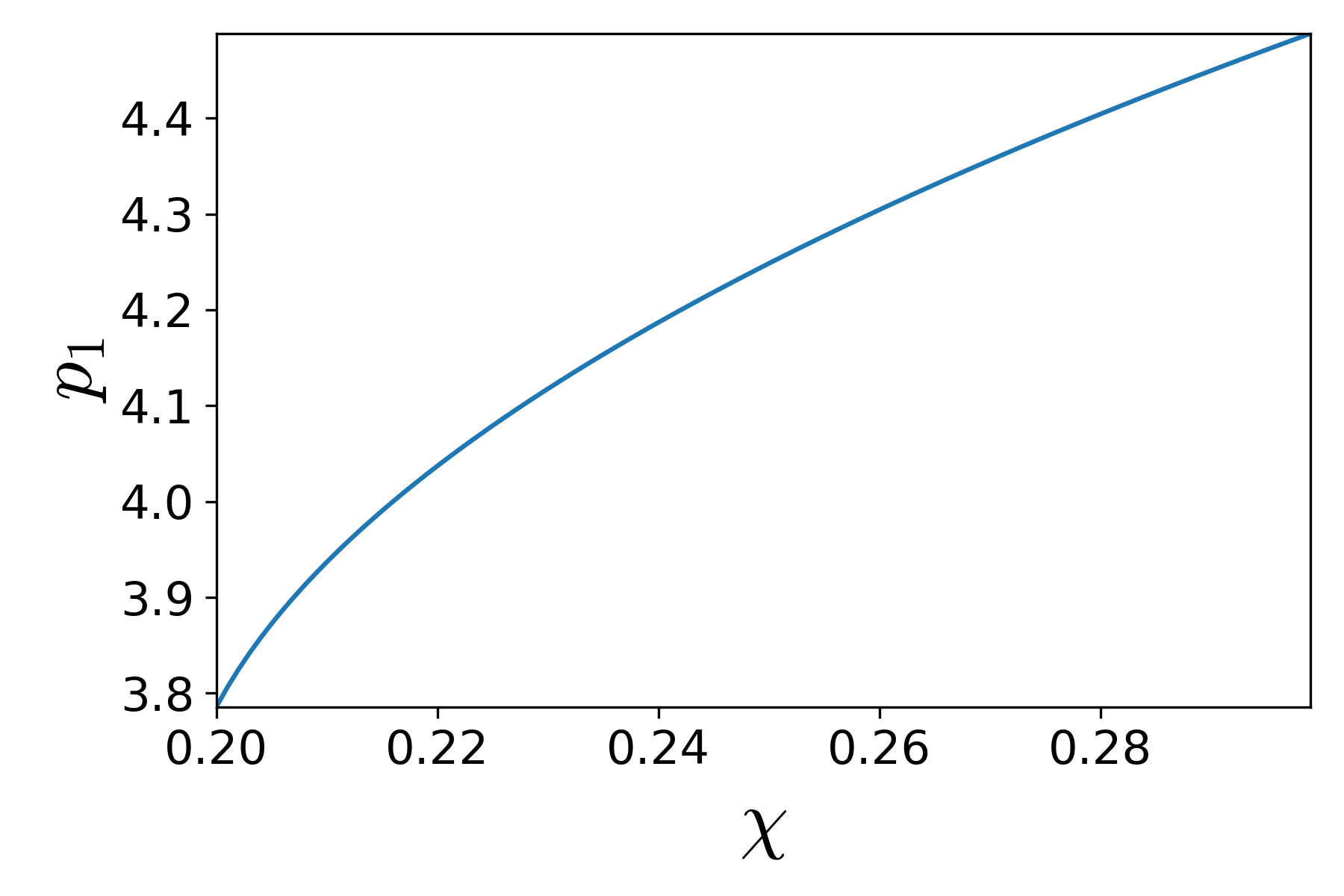}}
  \caption{Effective potentials with $E = 0.995\mu$, $L = 3.75365\mu M$ and $\zeta_\text{sGB} = 0.01$ for different values of the dimensionless spin parameter $\chi$. The innermost radius accessible to bound orbits, $p_1$, increases monotonically with $\chi$ (right). This implies a similar relationship as Fig. \ref{fig:veff_zeta}, but for spin rather than coupling. The same relationship also holds in dCS gravity.}
  \label{fig:veff_chi}
\end{figure*}

One may naively think that another way to amplify the effects of chaos would be to increase the size of the coupling constant $\zeta_\text{q}$, since after all this controls the magnitude of the GR deformation. This turns out not to be the case: as $\zeta_\text{q}$ is increased, the effective potential shifts away from the BH event horizon, (equivalent to saying that $p_1$ shifts away from the origin). Choosing an orbit with as small a $p_1$ as possible and a spacetime with as large a $\zeta_\text{q}$ as possible is therefore a difficult balance. We see this in \fig{fig:veff_zeta}, where in the left panel $p_1$ moves to larger values in $r/M$ as $\zeta_\text{q}$ is increased, and, on the right, that this increase is monotonic.  A similar relationship exists for the dimensionless spin parameter, $\chi$.  As seen in \fig{fig:veff_chi}, $p_1$ is pushed to larger values with increasing $\chi$.

We must now select a set of initial conditions to explore and construct a Poincaré section. To do so, given the choice of $E$ and $L$ discussed above, the second root of the effective potential serves as an initial guess about where to look for chaos.  We start by exploring a range of initial radii (for example, in \fig{fig:a_def}, the radius ranges from $(4.000, 4.005)$). We then initialize all of our geodesics with zero radial momenta, $P_r = 0$, and at the equator with $\theta = \pi/2$ and $\phi = 0$. Given this, we then obtain $P_\phi$ from the angular momentum $L$ definition, and solve for $P_\theta$ from the Hamiltonian (see Eq.~\eqref{eq-H-Veff}). Having completely defined the initial conditions of our geodesics, we are able to see if the rotation curve intersects the 1/2 resonance.  If it does not, we move the radius range until it does, and then gradually zoom-in until the value for $\mathcal{A}$ changes by less than 0.1\%.

\subsection{Numerical Implementation}

In order to search for chaos in geodesic motion, we first require a high-precision geodesic integrator that is able to capture equatorial surface crossings with the highest accuracy possible. To this end, we implemented a Runge-Kutta-style integrator due to Verner, of order (7,8) and adaptive \cite{hairer_solving_2009}. After explorations of the convergence properties of this integrator, we found it sufficient to set the integration tolerance to $10^{-15}$, the initial timestep to $10^{-1}$ and the smallest allowable timestep to $10^{-15}$. In order to capture surface crossings as accurately as possible, we implemented a bisection method every time an equatorial crossing was detected. We found it sufficient to set the tolerance in the bisection method to $10^{-15}$. With these choices, we are able to achieve machine precision in the calculation of the phase space position at each crossing.

This code was validated by performing much longer runs than required ($10^8$ orbits), and tracking the evolution of the constants of the motion.  In the process of validation, we found that some of the longer runs accumulated numerical error in the conserved quantities, which could lead to errors in the plateau size larger than the size of the plateaus we were looking for.  To mitigate this, we re-started the geodesic integrator every $10^4$ surface crossings, with initial conditions recalculated at the last surface crossing. More specifically, we re-initialized the initial conditions by setting $\theta = \pi/2$, and obtaining $P_\phi$ and $P_\theta$ from $L$, $E$ and the Hamiltonian; the values of $r, \phi$, and $P_r$ are copied from the previous integration point. This is allowed because, as stated previously, the surface crossing is a point in phase space that is known to the highest possible precision due to the bisection method.

A non-trivial aspect of the procedure laid out in the previous section is accurately determining the invariant point. Because the Poincaré surfaces can lie arbitrarily close to the invariant point, it can be difficult to balance finding the invariant point at the desired accuracy with minimizing the amount of computation time needed to find it.  We therefore adopt a centroid method to determine the invariant point as follows. As the initial condition approaches the invariant point, the shape of the Poincaré surfaces necessarily approaches a circle (the surfaces can be thought of as ``closing in'' on a single point  \cite{contopoulos_order_2002}).  Therefore, we track the area of the calculated Poincaré surface and compare it with the area of a hypothetical perfect circle, constructed from the largest and smallest $P_r$ values of the Poincar\'e surface. When these two areas are within $.1\%$ of each other, we then take the invariant point to be the center of the circle. In practice, this method requires approximately 5--6 iterations, using a bisection method in $r$ to refine the next guess, leading to an uncertainty in the measurement of the resulting plateau of $\delta \mathcal{A} \sim 10^{-14}$.

Finally we perform calibration runs in order to determine the numerical error inherent in the plateau calculation. We know that when the coupling parameter $\zeta_\text{q} = 0$, no plateaus should be present in the rotation curve. Due to numerical error, however, one may find an extremely small plateau. To determine the size of this numerical plateau, we set $\zeta_\text{q}$ to extremely small values on the order of $\zeta_\text{q} = 10^{-10}$ or smaller), and find a plateau of size $\mathcal{A}_0 \approx 10^{-12}$ irrespective of the small choice of $\zeta_\text{q}$. While we performed calibration runs for each new system we studied (for either geodesics in sGB gravity, dCS gravity, as well as for different choices of $(E, L)$ and of $\chi$), in practice the difference in the value of $\mathcal{A}_0$ was not appreciable between these systems. The plateau area $\mathcal{A}_0$ is therefore a lower bound on the smallest resolvable plateau of our implementation. Henceforth, we will compare any measured plateau for any finite value of $\zeta_\text{q}$ to $\mathcal{A}_0$ by computing $\mathcal{A}/\mathcal{A}_0$, and if $\mathcal{A}/\mathcal{A}_0 > 1$, then the chaotic features present in $\mathcal{A}$ can be interpreted as not arising from numerical error. 

We can now appreciate how computationally intensive it is to find a single plateau area.  Each geodesic must be integrated for about $10^7$ timesteps, and each plateau requires a couple hundred geodesics at the final range in $r$ (not to mention the low-resolution runs of several dozen geodesics each at three to four different zoom-in levels). Then, when this final set of geodesics have been integrated, calculating the rotation numbers and plateau area is a memory-intensive process; this is because calculating the rotation number in a reasonable amount of time requires storing as much phase-space data as possible in memory. For this reason, we parallelize the integration and deploy it on a high-performance computing cluster, dedicating one core to each geodesic.  When considering the time required for the calculation of a single plateau area, a single data point in, for example, \fig{fig:a_comp} requires about 500 CPU hours to obtain.

\subsection{Searching for chaos in quadratic gravity}

The bulk of our analysis focused on sGB gravity as the test case, and so we will focus mostly on this theory, but we repeated all calculations for dCS gravity and found very similar conclusions. The only significant difference between the two theories in our calculations is the size of the coupling parameters $\zeta_\text{q}$ that will generate chaotic features.

In performing this analysis, we are interested in capturing the effect on the plateau size from three free parameters, namely
\begin{enumerate}
    \item[(1)] {\textit{The expansion order in the spin of the GR sector}}, i.e.,~the coefficient $n$ in  $\mathcal{O}(\chi^n_\text{GR}, (\chi^m \zeta)_\text{q})$.  We expect any signatures of chaos induced by the truncation of the slow-rotating Kerr background to diminish as $n$ is increased, because an exact Kerr geometry does not permit chaos.
    \item[(2)] {\textit{The expansion order in the spin of the quadratic sector}}, i.e.,~the coefficient $m$ in $\mathcal{O}(\chi^n_\text{GR}, (\chi^m \zeta)_\text{q})$.  Varying this parameter will determine whether any chaotic signatures we find are induced by the modified theory or from the truncation of the slow-rotating expansion of the metric deformation. 
    \item[(3)] {\textit{The modified theory, coupling parameter}}, $\zeta_\text{q}$.  Varying the coupling will determine the sensitivity of chaotic signatures to the modified gravity deformations.
\end{enumerate}

\begin{figure*}[t]
    \centering
    \includegraphics[scale=0.57, clip=true]{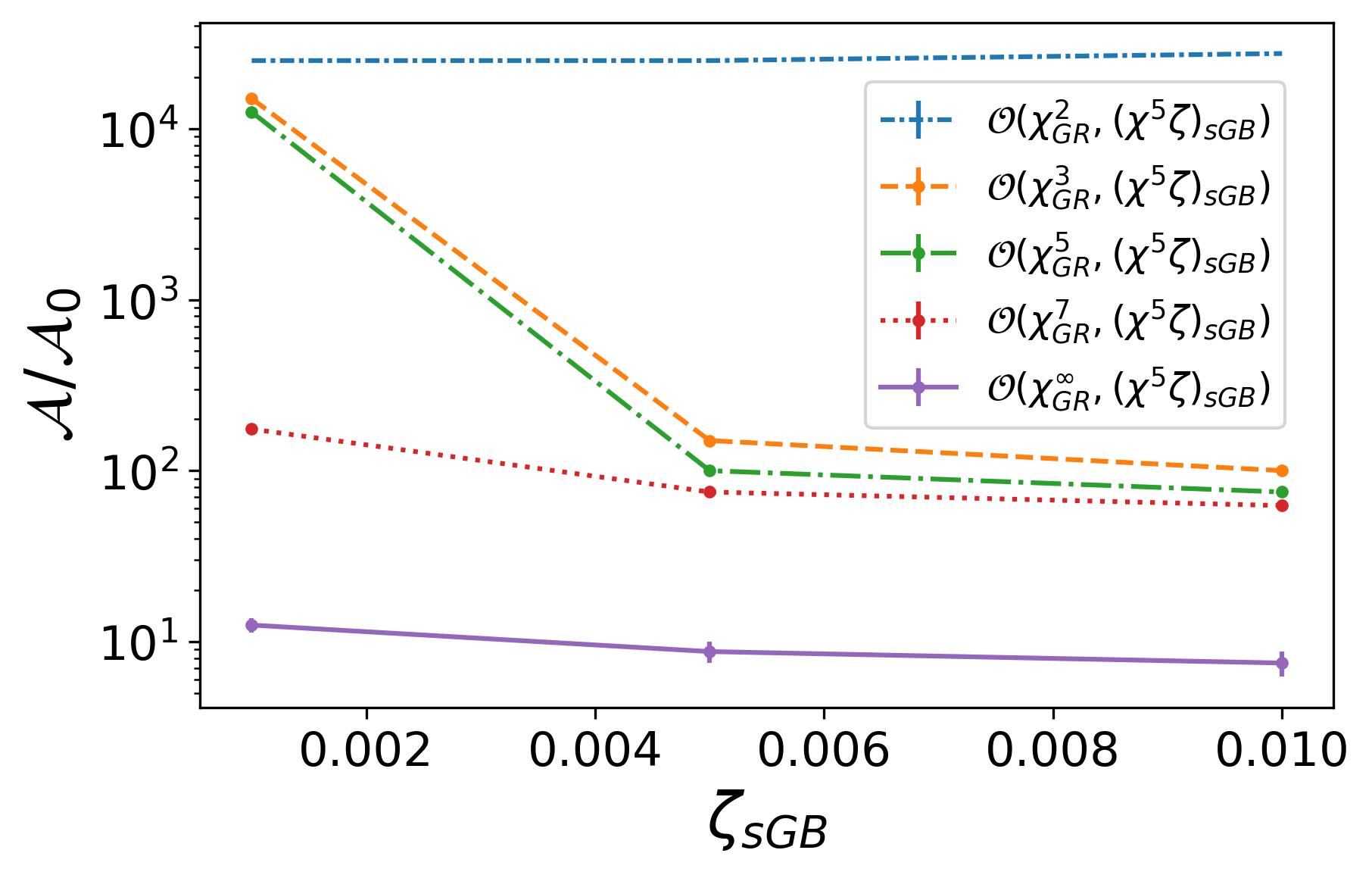}
    \includegraphics[scale=0.57, clip=true]{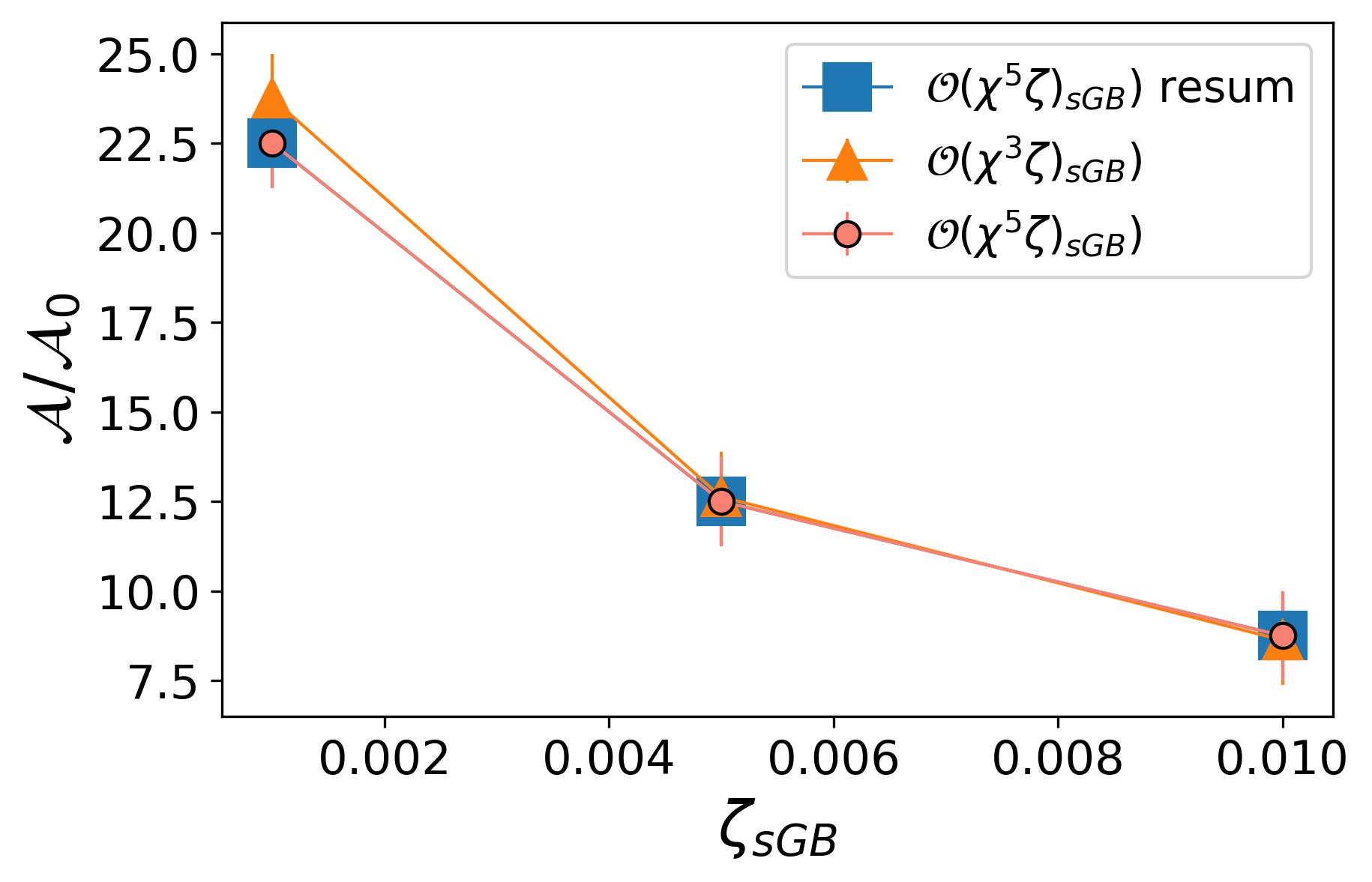}
    \caption{Left: Plateau areas for a system with an sGB modification, and orbital parameters $E=0.995\mu, L=3.75365\mu M$ and $\chi=0.2$.  The saturation of the size of the plateau as the expansion around $\chi$ increases indicates that chaos is present in geodesic trajectories, and suggests spinning BH solutions in this theory do not possess the same number of constants of the motion as a Kerr background. Right: Plateau areas for an sGB system with parameters $E=0.97406\mu, L=3.49916\mu M$ and $\chi=0.3$. The curves show that, despite different expansion orders in the sGB metric deformation sector, the plateau areas are identical, to within the precision of the integration.}
    \label{fig:a_comp}
\end{figure*}

Let us first consider varying $n$ and $\zeta_\text{q}$, as explored in the left panel of \fig{fig:a_comp}.  The blue line at the top shows that the largest signature of chaos occurs for the lowest value of $n$.  This is to be expected; any signature of chaos due to a quadratic gravity modification and not due to truncation of the slow-rotating approximation is here completely dominated by the latter.  As $n$ is increased, the plateau areas decrease, until they saturate to areas of about $\mathcal{A}/\mathcal{A}_0 \approx 10^1$ when $n \to \infty$. This $n \to \infty$ case corresponds to taking the GR part of the metric to be the Kerr metric exactly~\cite{contopoulos_order_2002}. These results imply that the chaotic signatures we find arise from the quadratic gravity modifications to the metric, and not from numerical error or from the truncation of the slow-rotation expansion of the GR sector of the metric.

As a side note, the left panel of \fig{fig:a_comp} also shows how chaotic signatures scale with $\zeta_\text{q}$. Focusing on the $n \to \infty$ curve, the general trend, to the extent there is one, is that the plateau size \emph{shrinks} with larger $\zeta_\text{q}$. As explained in Sec.~\ref{sec:theory}, this occurs because as the coupling constant is increased, the inner ``wall'' of $V_\text{eff}$ is pushed to larger radii.  This, in turn, means that bound orbits for larger values of $\zeta_\text{q}$ are forced to explore regions of phase space that are farther from the central body, and as such, are pushed farther away from the strong field regime.  Therefore, any subtle effects induced by high-curvature corrections near the event horizon are suppressed as $\zeta_\text{q}$ increases.  

These chaotic signatures, however, could be arising from the truncation of the slow-rotation expansion of the GR deformations of the metric, which we investigate by varying the order parameter $m$ in the right panel of \fig{fig:a_comp}. This figure shows that two metrics with deformations calculated to different orders in spin present plateaus that are identical to within numerical precision. The error bars in this figure are determined by taking each parameter that can affect the calculation of $\mathcal{A}$ (integration time, calculation of the invariant point, and resolution in $r$) to an order of magnitude greater precision that what was used in the runs and using the resulting difference in the plateau area as the error in the measurement of $\mathcal{A}$.

One can also repeat the calculation with a metric that restores higher order terms that correct the locations of the ergosphere and event horizon, as shown explicitly in the appendix of~\cite{2022}. This resummation is designed to recover the exact Kerr metric when given the Kerr metric expanded in $\chi$ to a finite order.  We see from \fig{fig:a_comp} that the plateau area behaves the same even when we use this resumed metric. We conclude then that the plateaus are a signature of chaos that is due to the metric deformation, rather than the particular expansion order used in the metric derivation.  All of the results discussed thus far are summarized in \tab{tab:summary}.

\begin{table}
\caption{A summary of the dynamics of the plateau size as a function of the relevant free parameters. The plateau area $\mathcal{A}_0$ is the lower bound on the smallest resolvable
plateau of our implementation.}
\begin{tabular}{lll}
\begin{tabular}[c]{@{}l@{}}As \rule{0.5cm}{0.15mm} is\\ increased...\end{tabular} & $\mathcal{A}/\mathcal{A}_0$...     & because...                                                                                                                           \\ \hline
$\mathcal{O}(\chi_\text{GR})$                                                                 & \begin{tabular}[c]{@{}l@{}}decreases,\\then saturates\end{tabular}      & \begin{tabular}[c]{@{}l@{}}The plateau area is initially\\ dominated by the low-$\chi$\\ expansion (\fig{fig:a_comp}, left)\end{tabular}                      \\ \hline
$\mathcal{O}(\chi\zeta_\text{q})$                                                                 & stays the same & \begin{tabular}[c]{@{}l@{}}The chaos is generated\\by the $\zeta_\text{q}$ coupling, not\\the low-$\chi$ expansion (\fig{fig:a_comp}, right)\end{tabular} \\ \hline
$\zeta_\text{q}$                                                           & \begin{tabular}[c]{@{}l@{}}decreases \end{tabular} & \begin{tabular}[c]{@{}l@{}}The effective potential is\\ pushed out of the strong-field\\ regime (\fig{fig:veff_zeta})\end{tabular}  
\end{tabular}
\label{tab:summary}

\end{table}

\begin{figure*}[t]
    \centering
    \includegraphics[scale=0.53, clip=true]{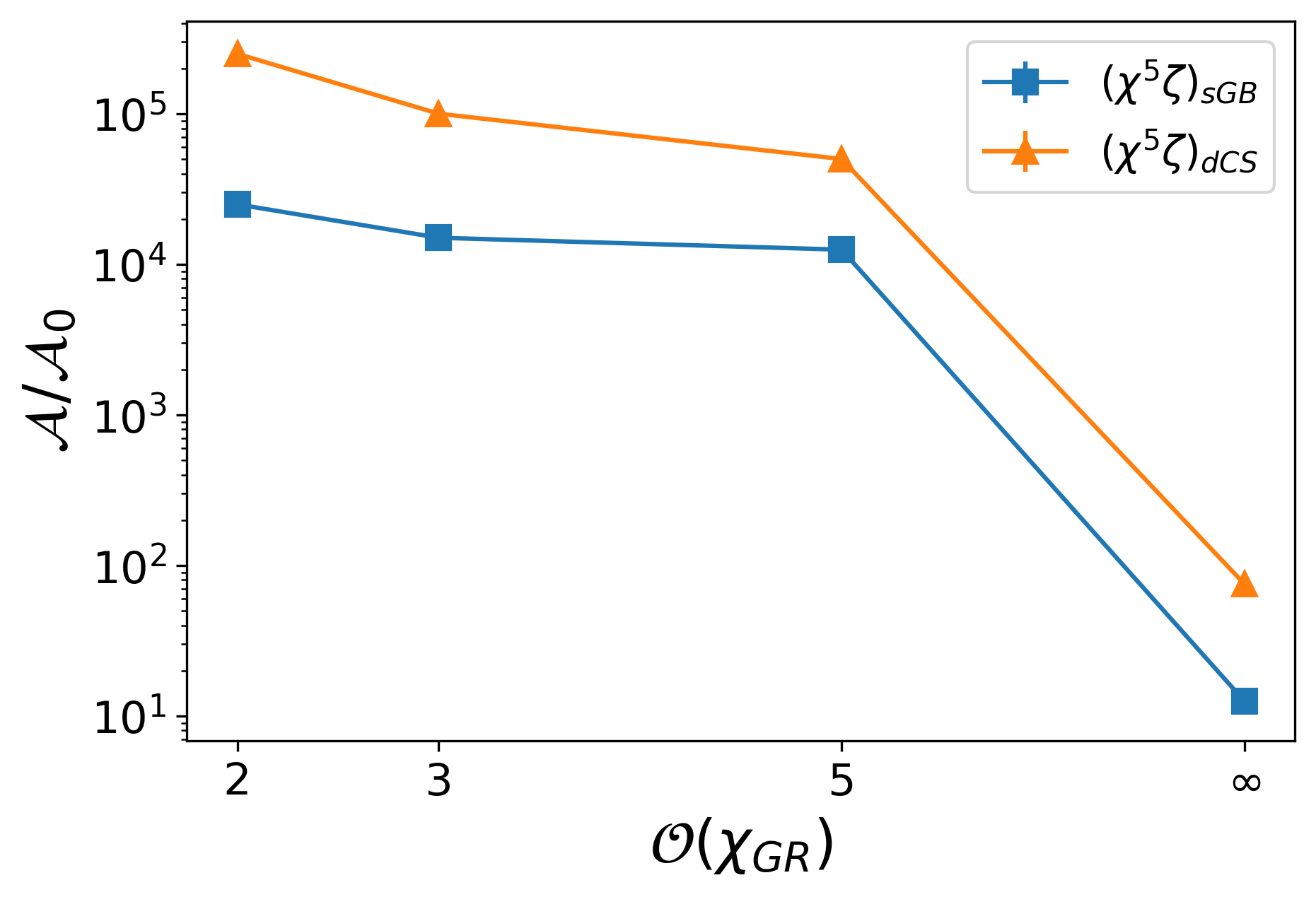}
    \includegraphics[scale=0.57, clip=true]{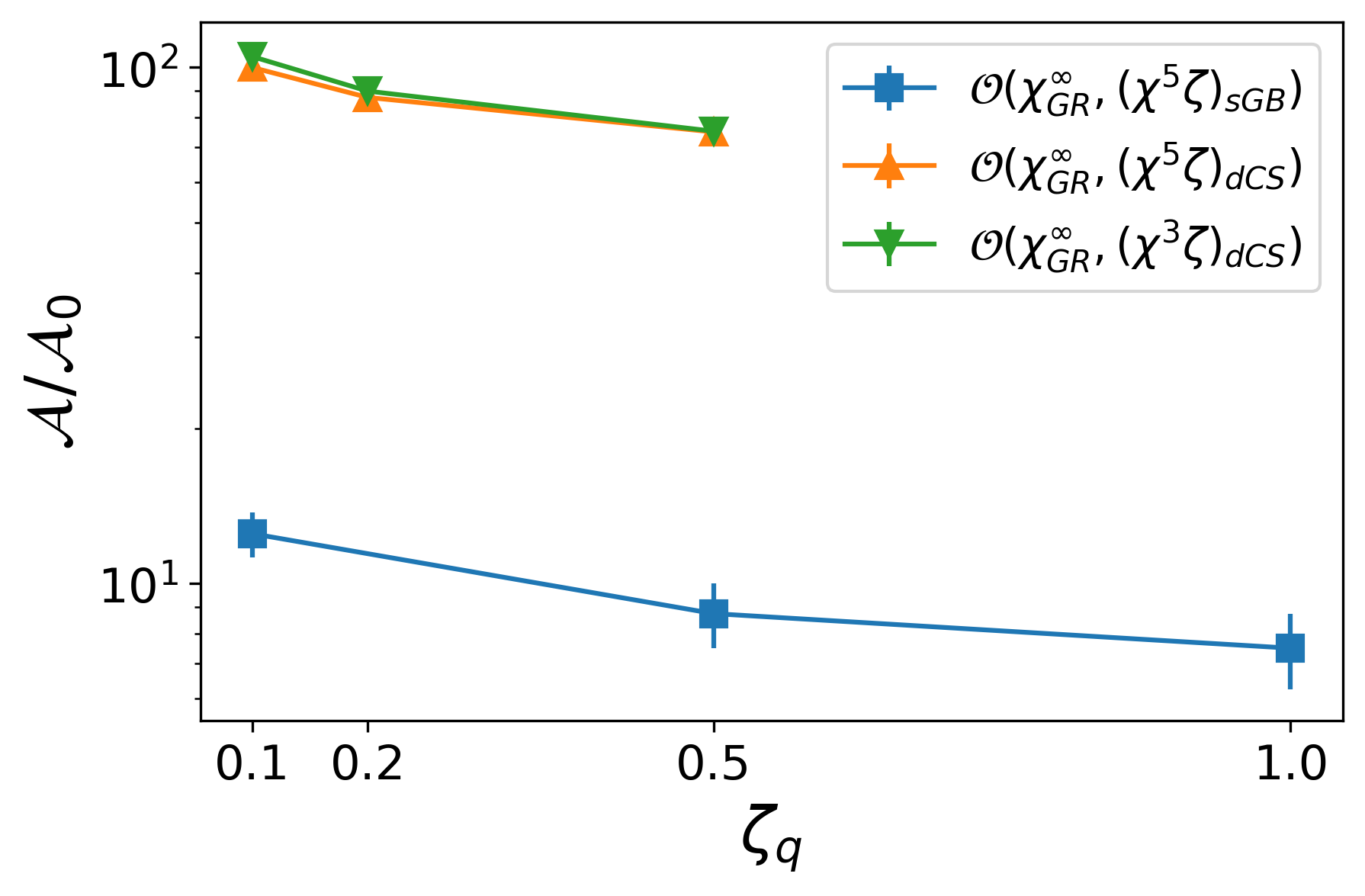}
    \caption{Left: Plateau areas as a function of truncation in slow-spin expansion in the GR sector, for $\zeta_\text{dCS} = 0.1$ and $\zeta_\text{sGB} = 0.001$.  Both theories show that as the GR sector approaches the Kerr metric, the chaotic features diminish but do not vanish, suggesting that any remaining chaos is due to the quadratic contribution. Right: Comparing plateau areas for geodesics in dCS and sGB gravity at $\mathcal{O}(\chi^\infty_{\text{GR}}, (\chi^5\zeta)_{\text{q}})$, with parameters $E=0.995\mu, L=3.75365\mu M, \chi=0.2$. The quantity $\zeta_\text{q} = \zeta_\text{dCS}$ in the dCS case, while $\zeta_\text{q} = 10^{2} \zeta_\text{sGB}$ in the sGB case. 
    }
    \label{fig:a_vs_n}
\end{figure*}

We have so far focused on sGB gravity, but we can repeat the entire analysis in dCS gravity. We obtain the same qualitative results when we use the dCS BH metric, as we now explain. When geodesics in dCS are computed at low-spin order in the GR sector, the chaos is overwhelmingly due to the slow-spin approximation.  However, as is the case in sGB gravity, when all spin orders are accounted for in the GR sector, there remains a small but measurable plateau, as shown in the left panel of \fig{fig:a_vs_n}.  When we compare the size of the plateaus in dCS gravity and sGB gravity, we find that the dCS plateaus are about an order of magnitude larger, as shown in the right panel of \fig{fig:a_vs_n}.  Furthermore, we confirm that when the metric deformation is truncated at low spin order, the plateau areas are indistinguishable in size from those obtained when the metric deformation is kept to higher order in spin.

%
%
%
\section{Discussion}

We performed an extensive numerical survey of geodesics in sGB and dCS gravity and found signatures of chaotic behavior in the geodesic motion independent of the perturbative scheme where these theories are derived and from numerical errors. These results imply the lack of a fourth constant of the motion or symmetry for rotating BHs in these theories. In order to study the robustness of the above conclusion, we studied whether the chaotic features remain when we consider resummed BH metrics, which formally contain an infinite number of spin terms. We found that the chaotic features do persist when using these metrics, and thus, our conclusions seem robust. One could argue that the still-unknown, exact (in spin), BH metric in these theories is sufficiently different from the resummed ones that chaos would disappear altogether if we had used such exact metrics. Although we have no evidence for this, one way to test this statement would be to perform a similar analysis on a numerically derived metric, such as those presented in \cite{sullivan_numerical_2020}. We expect our results to be true because numerical metrics (e.g.,~\cite{herdeiro2018asymptotically, sullivan_numerical_2020}) or particular extremal solutions (e.g.,~\cite{McNees:2015srl}) have been shown to lack features that are not already encapsulated by the slowly-rotating solutions. The re-analysis of chaos with numerically computed spacetimes would present several complications due to the numerical accuracy required for these calculations and add to the already considerable computational cost.  If the yet-to-be-known exact solution happens to possess a feature that is not already encapsulated by the slowly-rotating solutions, or that the behavior changes drastically, then our conclusions need to be revisited.

Let us now discuss more concretely the differences and similarities between this work and that of~\cite{cardenas-avendano_exact_2018}. While we did evolve the same exact geodesic parameters (BH spin and $\zeta_\text{q}$ parameters, as well as $E$ and $L$) and reproduced the results of~\cite{cardenas-avendano_exact_2018}, we were also able to explore many more values of $\zeta_\text{q}$ due to our new numerical implementation. Moreover, we defined a new measure, the plateau area, which enabled us to compare chaotic features with widely varying aspect ratios. This new measure, combined with our numerical accuracy, meant we could quantitatively study features in the rotation curve that previously appeared as discontinuous kinks, but are now revealed to have a width in $r/M$. Furthermore, we determined why areas become larger as $\zeta_\text{q}$ is increased, through an analysis of the roots of the effective potential.  All of this then led to the discovery that, although the size of the plateau areas do \textit{initially} decreases with truncation order of the deformed metric as found in~\cite{cardenas-avendano_exact_2018}, the size asymptotes to a constant; further evidence of this was then also found through the calculation of plateau areas using a resummed-in-spin metric deformation. All of this implies that, although the numerical calculations of~\cite{cardenas-avendano_exact_2018} are correct, their conjecture may not and chaos should be present in the full BH metric.

Given the characteristic size of the chaotic features we found, our results suggest that searching for chaos in gravitational-wave data from future detectors, (as suggested in Refs.~\cite{Babak:2006uv,apostolatos2009,Destounis:2021mqv}), may not be a viable method of placing constraints on the coupling parameters of these quadratic gravity theories. The largest chaotic signatures we found are truly tiny (around $10^{-5}$ in $r/M$), and they appear very close to the SMBH event horizon. This is in stark contrast to chaotic features that are found in other parametric BH spacetimes, which typically contain closed timelike curves or naked singularities. Chaotic features from well-motivated BH modifications may be significantly smaller than those in previously considered spacetimes, and thus doubt may be cast as to whether such features can be realistically detected. Exactly what the magnitude of these effects will be and what signal-to-noise ratio would be required to detect them necessitates significant work. What we have shown here contributes to the foundation for such work, as well as for the detailed study of EMRIs in quadratic theories of gravity.       

%
%
%
%

\section*{Acknowledgements} \label{sec:acknowledgements}

We thank Dimitry Ayzenberg, Kyriakos Destounis, Caroline Owen, Andrew Sullivan, and Yiqi Xie for useful comments and suggestions. A.~D.~and N.Y.~ acknowledge support from NASA ATP grant No. 17-ATP17-0225 and the Simons Foundation, Award number 896696. A.C.-A. acknowledges funding from the Fundaci\'on Universitaria Konrad Lorenz (Project 5INV1). Computations were performed on the Illinois Campus Cluster, a computing resource operated by the Illinois Campus Cluster Program (ICCP) in conjunction with the National Center for Supercomputing Applications (NCSA), which is supported by funds from the University of Illinois at Urbana-Champaign.

\appendix
%
%
%
%

\section{Liouville Integrability of arbitrary spherically symmetric stationary perturbations to Schwarzschild}

In this appendix, we provide a proof of the integrability of equations of motion, derived from a Schwarzschild metric with a spherically symmetric perturbation.  Consider the following metric:
\begin{equation*}
    \llmet = \llmet^{\mathrm{Sch.}} + \delta \llmet,
\end{equation*}
with $\delta \llmet$ defined (in Schwarzschild coordinates) as
\begin{equation}
\begin{split}
\delta ds^2 \equiv &\delta \llmet[g][tt](r) dt^2 + \delta \llmet[g][rr](r) dr^2 \\
&+r^2[\delta \llmet[g][\theta\theta](r) + \sin^2\theta\delta\llmet[g][\phi\phi](r)]\,.
\end{split}
\end{equation}
We then have for $\llmet$
\begin{eqnarray}
    \llmet[g][tt] &= -\left(1- \frac{2M}{r}\right) + \delta \llmet[g][tt](r),\\
    \llmet[g][rr] &= \left(1- \frac{2M}{r}\right)^{-1} + \delta \llmet[g][rr](r),\\
    \llmet[g][\theta\theta] &= r^2(1 + \delta \llmet[g][\theta\theta](r)),\\
    \llmet[g][\phi\phi] &= r^2\sin^2\theta(1 + \delta \llmet[g][\phi\phi](r))/,.
\end{eqnarray}
The angular components of this metric are equivalent to the angular components of the 2-sphere metric, except for a radially-dependent function (i.e. $r^2(1 + \delta \llmet (r))$), so the spherical symmetry is manifestly maintained. We could simplify this metric further by removing $\delta g_{\phi \phi}$ with a radial coordinate transformation, but, as we will show below, this is not necessary for the proof. 

Due to the spherical symmetry of the spacetime, without loss of generality, we can set $\theta = \pi/2$, leaving geodesic motion with 3 degrees of freedom: $t, r, \phi$.  Due to the stationarity and spherical symmetry of the spacetime, we also have two Killing vectors, given by $\partial_t$ (from time translation symmetry) and $\partial_\phi$ (from rotational symmetry about the $z$-axis).  Finally, given that the Hamiltonian $H = 1/(2\mu)\uumet p_\mu p_\nu$ is conserved (because the norm of the four-velocity is conserved), we now have  3 conserved quantities for the three degrees of freedom of geodesic motion. Therefore, by the Liouville-Arnold theorem, it is possible to transform the Hamiltonian to action-angle coordinates, and thus, the system is integrable and chaotic orbits are not permitted~\cite{contopoulos_order_2002}.

Explicitly, the perturbed Hamiltonian takes the form (for energy $E$ and $z$-component of angular momentum $L$)
\begin{eqnarray}
    H = &\frac{1}{2\mu}& \bigg[\frac{E^2}{\llmet[\delta g][tt](r)+\frac{2M}{r} - 1}
    +\frac{L^2}{\delta g_{\phi\phi}(r)+1}\\\nonumber
    &+&\left(\frac{p_\theta^2}{\delta g_{\theta\theta}(r) + 1}\right)r^{-2}
    + \frac{p_r^2}{\frac{2M}{r-2M} + \llmet[\delta g][rr](r) + 1}\bigg].
\end{eqnarray}  
Following through with Hamilton's formalism, we arrive at the equations of motion
\begin{align}
\dot{r} &= \frac{p_r}{\mu\left(1 + \frac{2M}{r-2M} + \llmet[\delta g][rr](r)\right)},\\
\dot{\theta} &= \frac{p_\theta}{\mu r^2(1 + \llmet[\delta g][\theta\theta](r))},\\
\dot{\phi} &= \frac{L}{\mu r^2(1 + \llmet[\delta g][\phi\phi](r))},\\
\dot{p}_r &= -\frac{1}{2\mu}\Bigg(-\frac{2\left(\frac{p_\theta^2}{1+\llmet[\delta g][\theta\theta](r)} + \frac{L^2}{1+\llmet[\delta g][\phi\phi](r)}\right)}{r^3}\\\nonumber
&{}\quad+\frac{p_r^2\left(2M-(r-2M)^2\llmet[\delta g][rr](r)'\right)}{r + (r - 2M)\llmet[\delta g][rr](r)^2}\\
&\quad+\nonumber
\frac{E^2(2M-r^2\llmet[\delta g][tt](r)')}{2M+r\llmet[\delta g][tt](r)^2}\\\nonumber
&{}\quad-\frac{\frac{p_\theta^2\llmet[\delta g][\theta\theta](r)'}{(1 + \llmet[\delta g][\theta\theta](r))^2} + \frac{L^2\llmet[\delta g][\phi\phi](r)'}{(1 + \llmet[\delta g][\phi\phi](r))^2}}{r^2}\Bigg),\\
\dot{p}_\theta &= \dot{p}_\phi = 0.
\end{align}
where primes indicate derivatives with respect to $r$.  Thus, the equations of motion can be written as first integrals of the motion. Note that this result does not depend on the perturbed metric being Schwarzschild; any spherically symmetric metric will do (of course, in GR, this will always be Schwarzschild).  Further, this says nothing about the \emph{size} of the perturbation $\delta g_{\mu\nu}$; it is a generic statement about the perturbation's \emph{symmetries}.

\bibliography{bib.bib}

\end{document}